\documentclass[preprintnumbers,amsmath,amssymb,prd,superscriptaddress]{revtex4}

\usepackage{latexsym}
\usepackage{amssymb}
\usepackage{epsfig,amsmath,graphics}
\usepackage{epstopdf}
\usepackage{verbatim}
\usepackage{wasysym}
\usepackage{hyperref}
\usepackage{feynmp-auto} 
\usepackage[utf8]{inputenc}
\usepackage{xpatch}
\usepackage{xcolor}
\usepackage{mathtools}
\hypersetup{
    colorlinks,
    linkcolor={red!80!black},
    citecolor={green!60!black},
    urlcolor={blue!60!black}}
\usepackage{appendix}

\newcommand{\Ez}{\mathcal{E}_0}
\newcommand{\Eboom}{\mathcal{E}_\text{boom}}
\newcommand{\OO}{\mathcal{O}}

\newcommand{\GeV}{\text{GeV}}
\newcommand{\MeV}{\text{MeV}}
\newcommand{\keV}{\text{keV}}

\newcommand{\cm}{\text{cm}}
\newcommand{\bn}{\text{b}} 
\newcommand{\mbn}{\text{mb}} 

\newcommand{\x}[1]{\ensuremath{\text{#1}}} 
\newcommand{\kin}{\text{kin}}
\newcommand{\xmin}{\text{min}}
\newcommand{\xmax}{\text{max}}
\newcommand{\ion}{\text{ion}}
\newcommand{\TF}{\text{TF}}
\newcommand{\LPM}{\text{LPM}}
\newcommand{\el}{\text{el}}
\newcommand{\inel}{\text{inel}}
\def\r{\right)}
\def\l{\left(}

\newcommand{\overbar}[1]{\mkern 1.5mu\overline{\mkern-1.5mu#1\mkern-1.5mu}\mkern 1.5mu}

\begin{document}


\title{White Dwarfs as Dark Matter Detectors}

\author{Peter W. Graham}
\affiliation{Stanford Institute for Theoretical Physics, Department of Physics,
Stanford University, Stanford, CA, 94305}

\author{Ryan Janish}
\affiliation{Berkeley Center for Theoretical Physics, Department of Physics,
University of California, Berkeley, CA 94720, USA}

\author{Vijay Narayan}
\affiliation{Berkeley Center for Theoretical Physics, Department of Physics,
University of California, Berkeley, CA 94720, USA}

\author{Surjeet Rajendran}
\affiliation{Berkeley Center for Theoretical Physics, Department of Physics,
University of California, Berkeley, CA 94720, USA}

\author{Paul Riggins}
\affiliation{Berkeley Center for Theoretical Physics, Department of Physics,
University of California, Berkeley, CA 94720, USA}

\begin{abstract}
Dark matter that is capable of sufficiently heating a local region in a white dwarf will trigger runaway fusion and ignite a type Ia supernova.
This was originally proposed by Graham et al. and used to constrain primordial black holes which transit and heat a white dwarf via dynamical friction.
In this paper, we consider dark matter (DM) candidates that heat through the production of high-energy standard model (SM) particles, and show that such particles will efficiently thermalize the white dwarf medium and ignite supernovae.
Based on the existence of long-lived white dwarfs and the observed supernovae rate, we derive new constraints on ultra-heavy DM with masses greater than $10^{16} ~\GeV$ which produce SM particles through DM-DM annihilations, DM decays, and DM-SM scattering interactions in the stellar medium. 
As a concrete example, we place bounds on supersymmetric Q-ball DM in parameter space complementary to terrestrial bounds.
We put further constraints on DM that is captured by white dwarfs, considering the formation and self-gravitational collapse of a DM core which heats the star via decays and annihilations within the core.
It is also intriguing that the DM-induced ignition discussed in this work provide an alternative mechanism of triggering supernovae from sub-Chandrasekhar, non-binary progenitors.
\end{abstract}

\maketitle

\section{Introduction}
\label{sec:intro}
Identifying the nature of dark matter (DM) remains one of the clearest paths beyond the Standard Model (SM) and it is thus fruitful to study the observable signatures of any yet-allowed DM candidate.
Many direct detection experiments are designed to search for DM, e.g.~\cite{Akerib:2016vxi, Agnese:2017njq}, yet these lose sensitivity to heavier DM due to its diminished number density.
Even for a strongly-interacting candidate, if the DM mass is above $\sim 10^{22}~\GeV$ a terrestrial detector of size $\sim (100~\text{m})^2$ will register fewer than one event per year.
While these masses are large compared to those of fundamental particles, it is reasonable to suppose that DM may exist as composite states just as the SM produces complex structures with mass much larger than fundamental scales (e.g., you, dear reader).
Currently there is a wide range of unexplored parameter space for DM candidates less than $\sim 10^{48}~\GeV$, above which the DM will have observable gravitational microlensing effects~\cite{Griest:2013aaa}.
For such ultra-heavy DM, indirect signatures in astrophysical systems are a natural way forward.
One such signal first proposed in~\cite{Graham:2015apa} is that DM can trigger runaway fusion and ignite type Ia supernovae (SN) in sub-Chandrasekhar white dwarf (WD) stars.

In addition to constraining the properties of DM, this raises the intriguing possibility that DM-induced runaway fusion is responsible for a fraction of observed astrophysical transients.
The progenitors of type Ia SN are not fully understood~\cite{Maoz:2012}, and recent observations of sub-Chandrasekhar~\cite{Scalzo:2014sap, Scalzo:2014wxa}, hostless~\cite{McGee:2010}, and unusual type Ia SN~\cite{Foley:2013} suggest that multiple progenitor systems and ignition mechanisms are operative.
Other suspected WD thermonuclear events, such as the Ca-rich transients~\cite{Kasliwal:2012}, are also poorly understood.
While mechanisms for these events have been proposed~\cite{Woosley1994,Fink:2007fv,Pakmor:2013wia,Sell:2015rfa}, the situation is yet unclear and it is worthwhile to consider new sources of thermonuclear ignition.

Runaway thermonuclear fusion requires both a heating event and the lack of significant cooling which might quench the process.
The WD medium is particularly suited to this as it is dominated by degeneracy pressure and undergoes minimal thermal expansion, which is the mechanism that regulates fusion in main sequence stars.
Thermal diffusion is the primary cooling process in a WD, and it can be thwarted by heating a large enough region.
The properties of a localized heating necessary to trigger runaway fusion were computed in~\cite{Woosley}.
Consequently, it was realized~\cite{Graham:2015apa} that if DM is capable of sufficiently heating a WD in this manner, it will result in a SN with sub-Chandrasekhar mass progenitor.
This was used to place limits on primordial black holes which transit a WD and cause heating by dynamical friction, although the authors of~\cite{Graham:2015apa} identify several other heating mechanisms which may be similarly constrained. 
Note that the idea of using observations of WDs to constrain DM properties has been pursued before, e.g. through an anomalous heating of cold WDs~\cite{Bertone:2007ae, McCullough:2010ai} or a change in the equilibrium structure of WDs with DM cores~\cite{Leung:2013pra}. 
These are quite distinct from the observational signature considered in this work, which is the DM trigger of a type Ia SN (although see~\cite{Bramante:2015cua} for a related analysis).

In this paper, we examine DM candidates which have additional non-gravitational interactions and are thus capable of heating a WD and igniting a SN through the production of SM particles. 
An essential ingredient in this analysis is understanding the length scales over which SM particles deposit energy in a WD medium.
We find that most high energy particles thermalize rapidly, over distances shorter than or of order the critical size for fusion.
Particle production is thus an effective means of igniting WDs. 
Constraints on these DM candidates come from either observing specific, long-lived WDs or by comparing the measured rate of type Ia SN with that expected due to DM.
It is important to note that these constraints are complementary to direct searches---it is more massive DM that is likely to trigger SN, but also more massive DM that has low terrestrial flux.
The WD detector excels in this regime due to its large surface area $\sim (10^4~\text{km})^2$, long lifetime $\sim \text{Gyr}$, and high density.
We demonstrate these constraints for generic classes of DM models that produce SM particles via DM-SM scattering, DM-DM collisions, or DM decays, and consider the significantly enhanced constraints for DM that is captured in the star.
For these cases, we are able to place new bounds on DM interactions for masses greater than $m_\chi \gtrsim 10^{16}~\GeV$.
As a concrete example we consider ultra-heavy Q ball DM as found in supersymmetric extensions of the SM. 

The rest of the paper is organized as follows.
We begin in Section~\ref{sec:boomreview} by reviewing the mechanism of runaway fusion in a WD.
In Section~\ref{sec:smheating} we study the heating of a WD due to the production of high-energy SM particles.
Detailed calculations of the stopping of such particles are provided in Appendix~\ref{sec:wdpdg}.
In Section~\ref{sec:dmignition} we parameterize the explosiveness and event rate for generic classes of DM-WD encounters, and in Section~\ref{sec:constraints} we derive schematic constraints on such models.
The details of DM capture in a WD are reserved for Appendix~\ref{sec:capture}.
Finally we specialize to the case of Q-balls in Section~\ref{sec:qballs}, and conclude in Section~\ref{sec:discussion}.

\section{White Dwarf Runaway Fusion}
\label{sec:boomreview}
We first review the conditions for which a local energy deposition in a WD results in runaway fusion.
Any energy deposit will eventually heat ions within some localized region---parameterize this region by its linear size $L_0$, total kinetic energy $\Ez$ and typical temperature $T_0$.
These scales evolve in time, but it will be useful to describe a given heating event by their initial values.

The fate of a heated region is either a nonviolent diffusion of the excess energy across the star, or a runaway fusion chain-reaction that destroys the star.
The precise outcome depends on $L_0$, $\Ez$ and $T_0$.
There is a critical temperature $T_f$, set by the energy required for ions to overcome their mutual Coulomb barrier, above which fusion occurs.
For carbon burning, $T_f \sim \MeV$~\cite{Gasques:2005ar}.
Any heated region $T_0 > T_f$ will initially support fusion, although this is not sufficient for runaway as cooling processes may rapidly lower the temperature below $T_f$.
This cooling will not occur if the corresponding timescale is larger than the timescale at which fusion releases energy.
Cooling in a WD is dominated by thermal diffusion, and the diffusion time increases as the size of the heated region.
However, the timescale for heating due to fusion is independent of region size.
Thus, for a region at temperature $\gtrsim T_f$, there is a critical size above which the heated region does not cool but instead initiates runaway.
For a region at the critical fusion temperature $T_f$, we call this critical size the \emph{trigger size} $\lambda_T$.
The value of $\lambda_T$ is highly dependent on density, and in a WD is set by the thermal diffusivity of either photons or degenerate electrons.
This critical length scale has been computed numerically in~\cite{Woosley} for a narrow range of WD densities and analytically scaled for other WD masses in~\cite{Graham:2015apa}.
As in~\cite{Graham:2015apa}, we will restrict our attention to carbon-oxygen WDs in the upper mass range $\sim 0.85 - 1.4 ~M_{\astrosun}$ (these will yield the most stringent constraints on DM).
This corresponds to a central number density of ions $n_\text{ion} \sim 10^{30} - 10^{32} ~\cm^{-3}$ and a trigger size of $\lambda_T \sim 10^{-3} - 10^{-5} ~\text{cm}$.

\begin{figure}
\includegraphics[scale=.35]{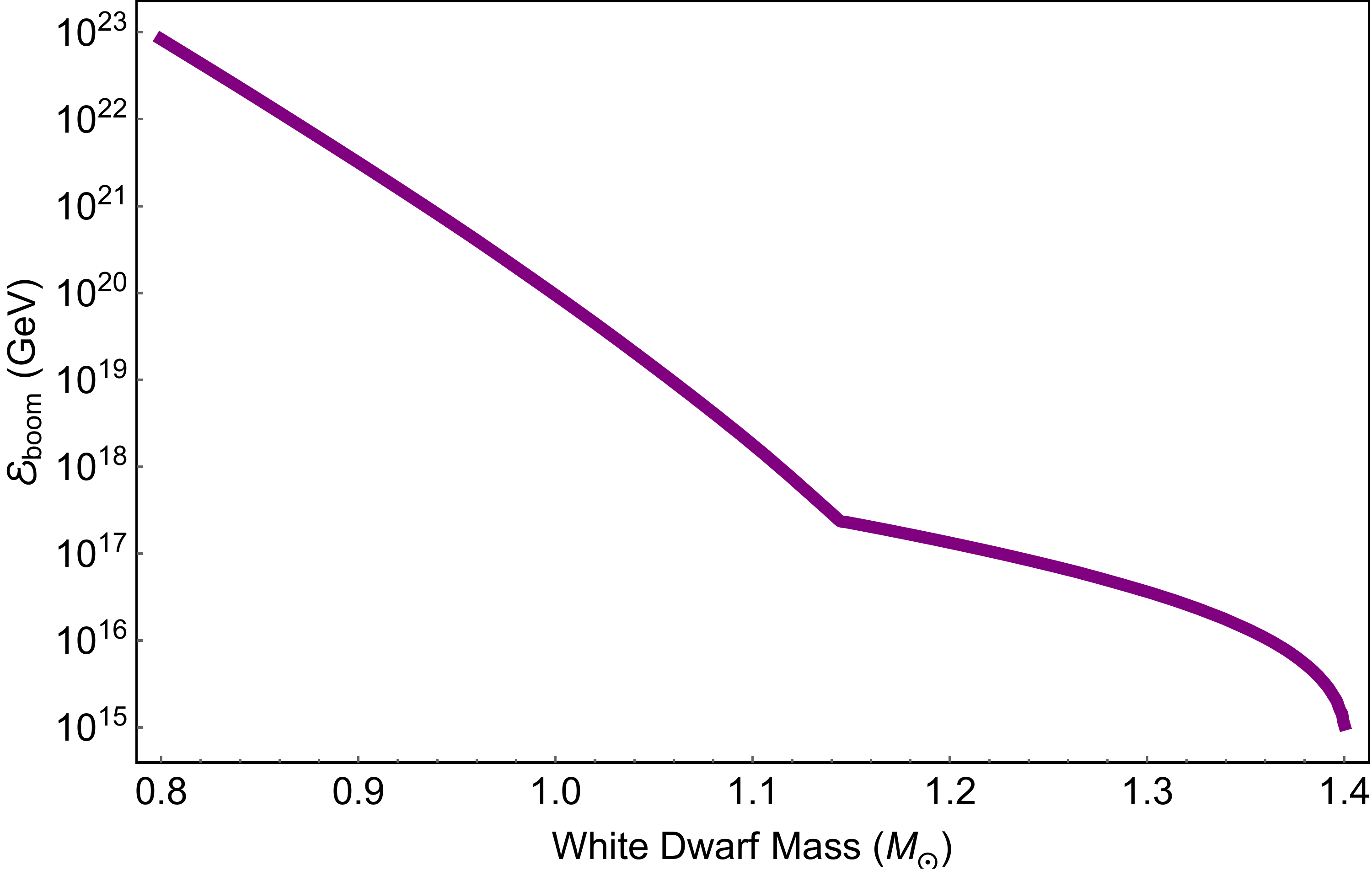}
\caption{The minimum energy deposit~\eqref{eq:Eboom} necessary to trigger runaway fusion, based on numerical results for $\lambda_T$~\cite{Woosley} and the WD mass-density relation~\cite{cococubed}}.
\label{fig:Eboom}
\end{figure}

If a heated region is smaller than the trigger size, its thermal evolution is initially dominated by diffusion.
However, this will still result in runaway fusion if the temperature is of order $T_f$ by the time the region diffuses out to the trigger size.
For our purposes it is more natural to phrase this in terms of the total energy $\Ez$ deposited during a heating event.
Of course, the relation between energy $\Ez$ and temperature $T_0$ depends on the rate at which WD constituents---ions, electrons, and photons---thermalize with each other within the region size $L_0$.
Given that the different species thermalize rapidly, the excess energy required to raise the temperature to $T_f$ in a volume $V$ is given by a sum of their heat capacities
\begin{equation}
\label{eq:heatcapacity}
  \frac{\Ez}{V} \gtrsim \int_0^{T_f} dT (n_\text{ion} + n_e^{2/3} T + T^3),
\end{equation}
where $n_e$ is the number density of electrons.
Note that we use the heat capacity of a degenerate gas of electrons, since the Fermi energy $E_F \gtrsim \MeV$ for the densities we consider.
The minimum energy deposit necessary to trigger runaway fusion is simply
\begin{align}
\label{eq:Eboom}
\Eboom &\sim \lambda_T^3 (n_\text{ion} T_f + n_e^{2/3} T_f^2 + T_f^4) \\
&\approx 10^{16} - 10^{23} ~\GeV. \nonumber
\end{align}
$\Eboom$ is shown over the range of WD masses in Figure~\ref{fig:Eboom}, where we have employed a numerical formulation of the WD mass-density relation as given by~\cite{cococubed}. 
Once again, for a given WD density the critical energy threshold is primarily set by $\lambda_T$---this length scale has been carefully computed and tabulated in~\cite{Woosley}, along with the attendant assumptions. 
In any case, we expect the simplified expression~\eqref{eq:Eboom} to be accurate at the order of magnitude level, and we refrain from a more detailed analysis here. 
Thus for a heating event characterized by its $L_0$, $\Ez$, and $T_0 \gtrsim T_f$, there is an \emph{ignition condition}:
\begin{align}
    \label{eq:energy_boom_condition}
    \Ez \gtrsim
    \Eboom \cdot \text{max}\left\{1, \frac{L_0}{\lambda_T}\right\}^3.
\end{align}
Any $\Ez$ satisfying this condition is minimized for $L_0$ less than the trigger size, where it is also independent of the precise value of $L_0$.
For broader deposits, the necessary energy is parametrically larger than $\Eboom$ by a volume ratio $(L_0/\lambda_T)^3$.
As a result, understanding the $L_0$ for different kinds of heating events in a WD is critical to determining whether or not they are capable of destroying the star.

\section{Particle Heating of White Dwarfs}
\label{sec:smheating}
Production of high-energy SM particles in a WD will result in heating of the stellar medium.
The critical quantity to understand is the length scale over which such heating occurs---this scale determines the efficiency of the heating event in triggering runaway fusion, as described by condition~\eqref{eq:energy_boom_condition}.
Note that this is a question of purely SM physics.
The unknown physics of DM will serve only to set the initial properties of the SM particles.

\begin{figure*}
\includegraphics[scale=1.0]{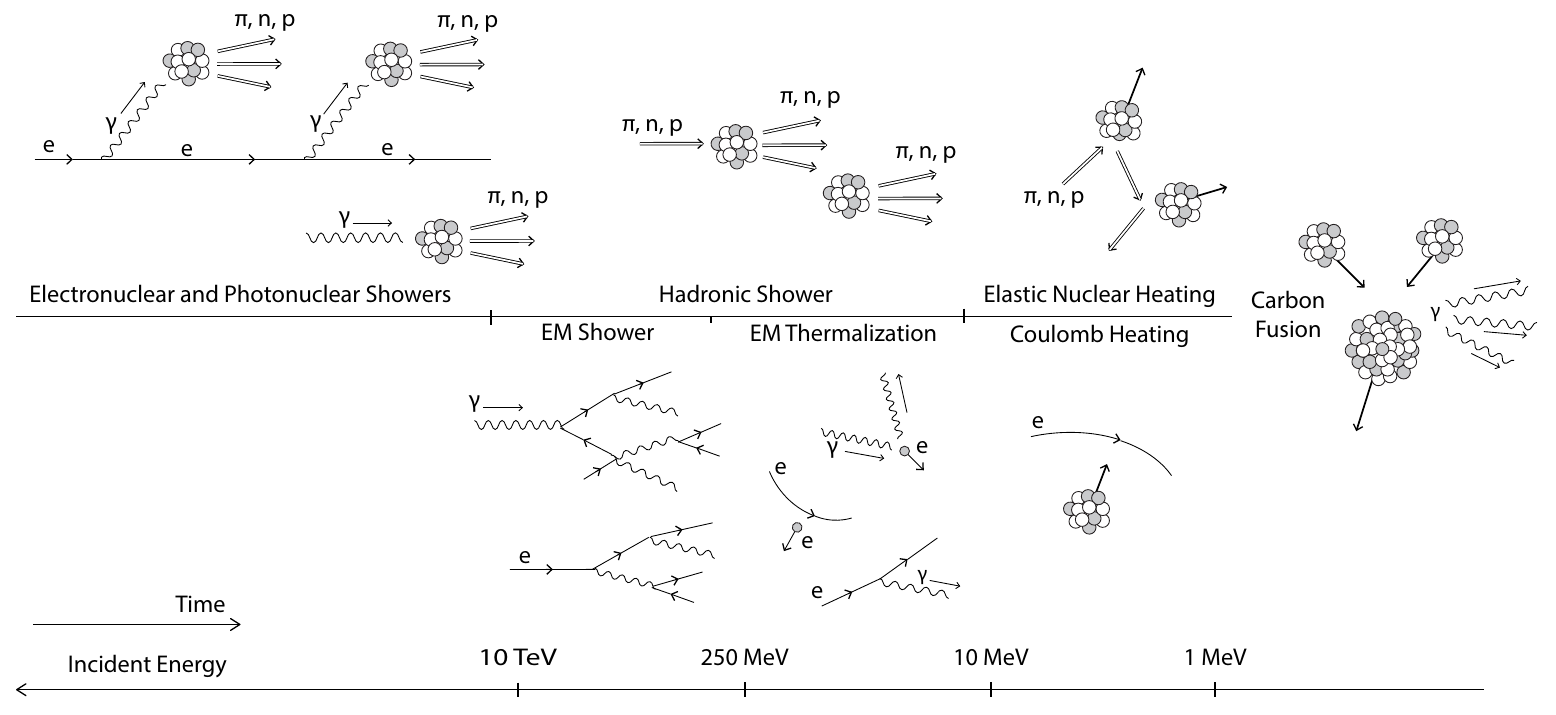}
\caption{Dominant energy loss and thermalization processes in the WD as a function of energy, with energy decreasing towards the right.
Hadronic processes are shown in the upper panel and EM processes in the lower panel.
High energy particles will induce showers that terminate into elastic thermalization of the WD ions, moving from left to right in the diagram.
The quoted energies are for a $\sim 1.37 ~M_{\astrosun}$ WD, although the cartoon is qualitatively the same for all densities.}
\label{fig:cooling-cartoon}
\end{figure*}

We find that SM particles efficiently heat the WD regardless of species or energy (neutrinos are a slight exception)---the heating length is typically less than or of order the trigger size $\lambda_T$.
This is accomplished primarily through hadronic showers initiated by collisions with carbon ions.
In some cases electromagnetic showers are important, however at high energies these are suppressed by density effects and even photons and electrons are dominated by hadronic interactions.
These showers rapidly stop high-energy particles due to their logarithmic nature, transferring the energy into a cloud of low-energy particles which heat the medium through elastic scatters.
A schematic for the flow of energy during deposition is given in Figure~\ref{fig:cooling-cartoon}.
In this light, the WD operates analogously to a particle detector, including hadronic and electromagnetic ``calorimeter'' components.
Runaway fusion provides the necessary amplification to convert a detected event into an observable signal.

The remainder of this section will discuss the above heating process in more detail.
We summarize the dominant source of energy loss and the resulting stopping lengths $\lambda$ for SM particles of incident kinetic energy $\epsilon$.
The total path length traveled by a particle before depositing $\OO(1)$ of its energy is approximately
\begin{equation}
R_\text{SP} \sim \frac{\epsilon}{dE/dx},
\end{equation}
where $dE/dx$ is the stopping power in the WD medium.
If the mean free path to hard scatter $\lambda_\text{hard}$ is smaller than this path length $R_\text{SP}$, then the particle undergoes a random walk with $N_\text{hard}$ scatters, and the net displacement is reduced by $\sqrt{N_\text{hard}}$.
We therefore approximate the stopping length as
\begin{align}
\lambda \sim \text{min}\left\{ R_\text{SP}, \sqrt{R_\text{SP}\lambda_\text{hard}} \right\}
\end{align}
This random walk behavior is relevant for low-energy elastic scatters.

Stopping lengths are plotted in Figures~\ref{fig:SPhighHad} and~\ref{fig:SPhighEM}, and a detailed treatment of the stopping powers is given in Appendix~\ref{sec:wdpdg}.
We will consider incident light hadrons, photons, electrons, and neutrinos---as we are concerned with triggering runaway fusion, we restrict our attention to energies $\epsilon \gg T_f \sim \text{MeV}$.

\subsection{High-Energy Showers}

\paragraph{Hadronic Showers.}
Incident hadrons with kinetic energy larger than the nuclear binding scale $\sim 10~\MeV$ will undergo violent inelastic collisions with carbon ions resulting in an $\OO(1)$ number of secondary hadrons.
This results in a roughly collinear shower of hadrons of size
\begin{align}
\label{eq:hadlength}
  X_\text{had} &\sim \frac{1}{n_\ion \sigma_\text{inel}} \log\l\frac{\epsilon}{10 ~\MeV}\r \\
  &\approx 10^{-6} ~\text{cm}
  \l\frac{10^{32}~\text{cm}^{-3}}{n_\text{ion}}\r. \nonumber
\end{align}
where the inelastic nuclear cross section is $\sigma_\text{inel} \approx 100 ~\text{mb}$ and we have taken the logarithm to be $\sim 10$.
The shower terminates into pions and nucleons of energy $\sim 10~\MeV$, whose cooling is discussed below.
Note that neutral pions of energy $10 - 100 ~\text{MeV}$ have a decay length to photons of $\delta_{\pi^0} \sim 10^{-6} ~\text{cm}$.
Hadronic showers will therefore generate an electromagnetic component carrying an $\OO(1)$ fraction of the energy.

\begin{figure}
\includegraphics[scale=.35]{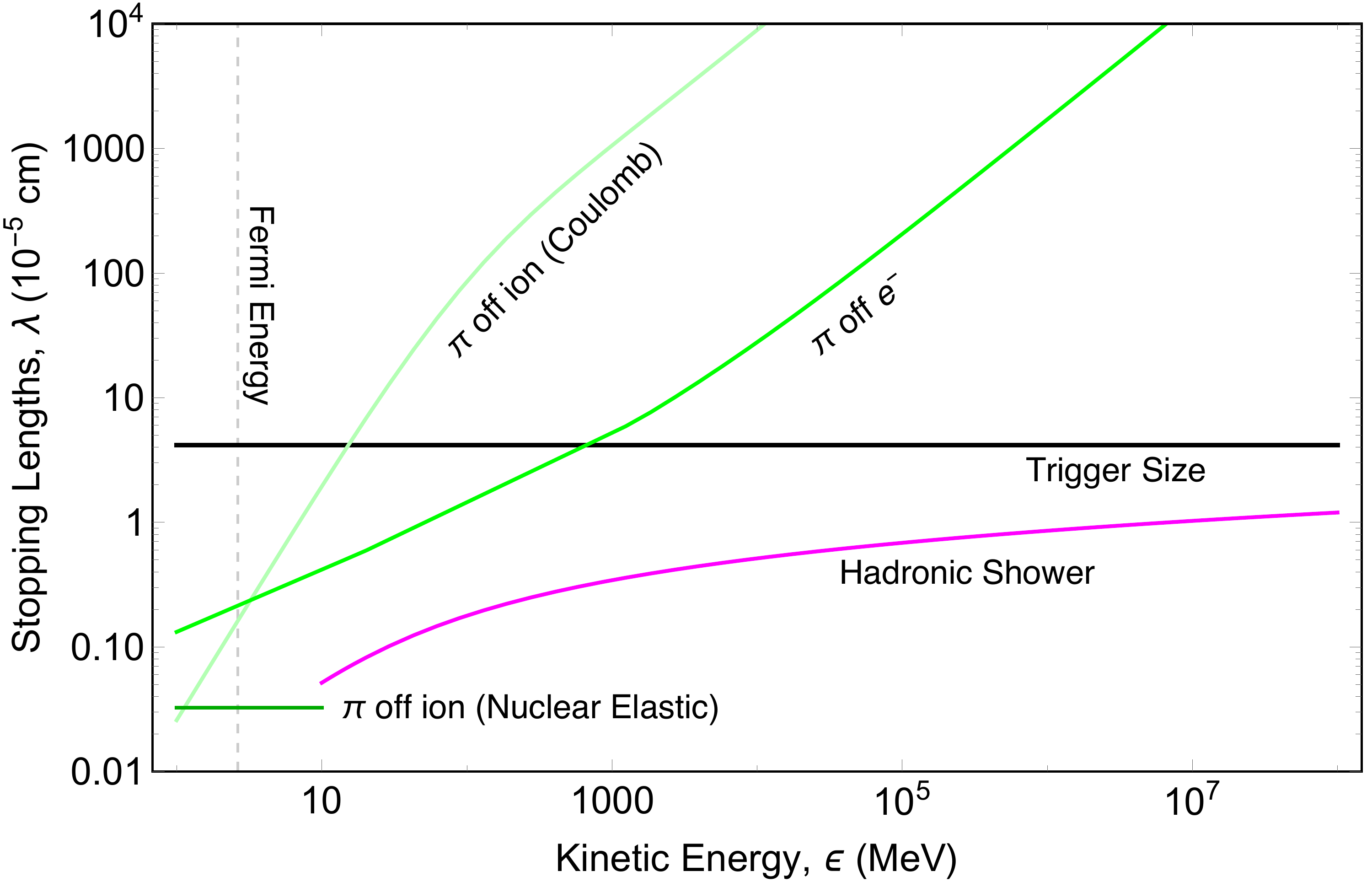}
\caption{Stopping lengths for incident hadrons as a function of kinetic energy in a WD of density $n_\text{ion} \sim 10^{31}~\text{cm}^{-3}$ ($\approx 1.25 ~M_{\astrosun}$), including the hadronic shower length (magenta).
Any discontinuities in the stopping lengths are due to approximate analytic results in the different energy regimes.
See Appendix~\ref{sec:wdpdg} for calculation details.
}
\label{fig:SPhighHad}
\end{figure}

\begin{figure}
\includegraphics[scale=.35]{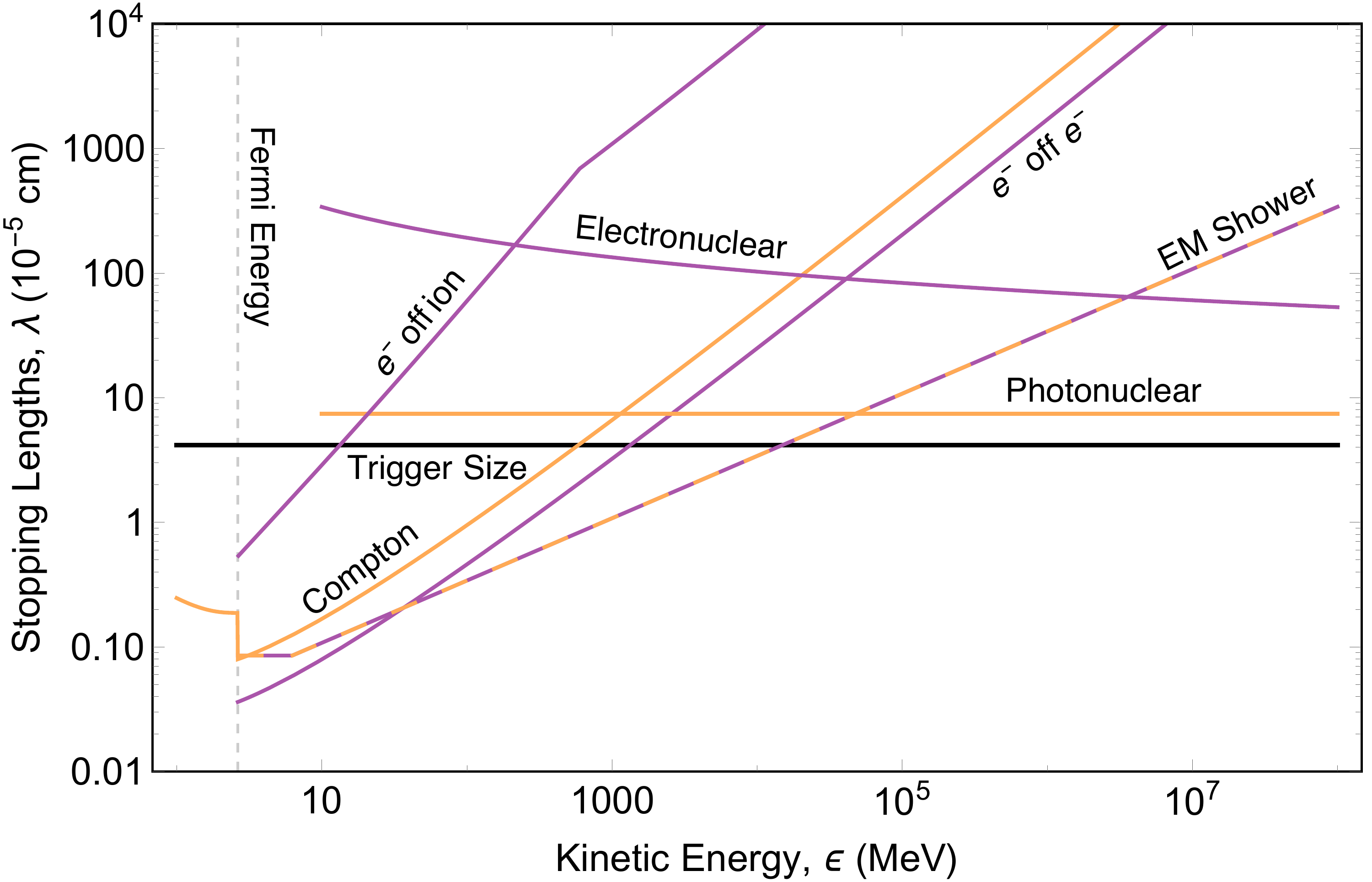}
\caption{Stopping lengths of incident photons (orange) and electrons (purple) as a function of kinetic energy in a WD of density $n_\text{ion} \sim 10^{31}~\text{cm}^{-3}$ ($\approx 1.25 ~M_{\astrosun}$), including the EM shower length (dashed).
Any discontinuities in the stopping lengths are due to approximate analytic results in the different energy regimes.
See Appendix~\ref{sec:wdpdg} for calculation details.
}
\label{fig:SPhighEM}
\end{figure}

\begin{figure}
\includegraphics[scale=1.0]{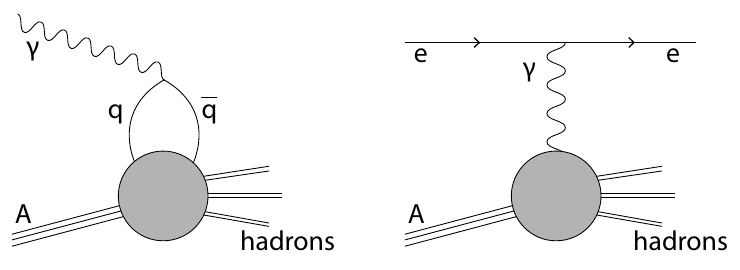}
\caption{Photonuclear (left) and Electronuclear (right) interactions. The shaded region contains, at high energies, the familiar point-like processes of deep inelastic scattering and for energies below $\Lambda_\x{QCD}$ is best described by exchange of virtual mesons.}
\label{fig:electrophotonuclear-diagram}
\end{figure}

\paragraph{Photonuclear and Electronuclear Showers.}
A photon or electron can directly induce hadronic showers via production of a quark-antiquark pair, depicted in Figure~\ref{fig:electrophotonuclear-diagram}.
The LPM effect, discussed below, ensures that these process dominate the stopping of photons and electrons at high energies, $\epsilon \gtrsim 10^4-10^6~\GeV$.

The only substantial difference between photonuclear showers and purely hadronic ones is that they require a longer distance to initiate.
Roughly, the photonuclear cross section is suppressed relative to the hadronic inelastic cross section $\sigma_\text{inel}$ by a factor of $\alpha$, and so the photon range is
\begin{align}
\label{eq:photonuclength}
  \lambda_{\gamma A} \approx 10^{-5} ~\text{cm} \l\frac{10^{32}~\text{cm}^{-3}}{n_\text{ion}}\r.
\end{align}
Here $\lambda_{\gamma A}$ is the distance to initiate a hadronic shower, whereas the shower itself extends a distance $X_\text{had}$.
Note that $\lambda_{\gamma A}$ is of order the trigger size.

The electronuclear showers are qualitatively different, as the electron survives the interaction.
This process is best described as a continuous energy loss of the electron, due to radiation of virtual photons into hadronic showers.
The stopping power is again radiative, which gives the constant stopping length
\begin{align}
\label{eq:electronuclength}
  \lambda_{eA}
  \approx 10^{-4} ~\text{cm} \l\frac{10^{32}~\text{cm}^{-3}}{n_\text{ion}}\r.
\end{align}
This is suppressed by an additional factor of $\alpha$ relative to the photonuclear interaction, although a full calculation also yields an $\OO(10)$ logarithmic enhancement.
We see that the electronuclear length scale $\lambda_{eA}$ is at most larger than the trigger size by an order of magnitude.

\paragraph{Electromagnetic Showers.}
Of course, electrons and photons can also shower through successive bremsstrahlung and pair-production.
An electromagnetic shower proceeds until a critical energy $\sim 100 ~\MeV$, at which point these radiative processes become subdominant to elastic Coulomb and Compton scattering.
Below this scale radiation can still be important, though electromagnetic showers do not occur.
Note that bremsstrahlung and pair-production are strictly forbidden for incident energies below the Fermi energy $E_F$.

At sufficiently high electron/photon energies and nuclear target densities, electromagnetic showers are elongated due to the Landau-Pomeranchuk-Migdal (LPM) effect.
High-energy radiative processes necessarily involve small momentum transfers to nuclei.
These soft virtual photons cannot be exchanged with only a single ion, but rather interact simultaneously with multiple ions.
This generates a decoherence, suppressing bremsstrahlung/pair-production above an energy $E_\text{LPM}$ which scales inversely with density:
\begin{align}
    E_\text{LPM} \approx 1~\MeV
    \l \frac{10^{32}~\text{cm}^{-3}}{n_\text{ion}} \r
\end{align}
The corresponding shower lengths are
\begin{align}
  X_\text{EM} &\approx X_0 \cdot
  \begin{cases}
  \l \frac{\epsilon}{E_\text{LPM}} \r^{1/2} & \epsilon > E_\text{LPM} \\
  \;\;\;\;\;\, 1 & \epsilon < E_\text{LPM}
  \end{cases}
\end{align}
where
\begin{align}
  X_0 &\approx 10^{-7} ~\text{cm}
  \l\frac{10^{32}~\text{cm}^{-3}}{n_\text{ion}}\r
\end{align}
is the unsuppressed EM shower length.
See Appendix~\ref{sec:emshowers} for details.
At the highest WD densities radiative processes are always LPM-suppressed, while at lower densities we observe both regimes.
We emphasize that for all densities, throughout the energy range where it is relevant, the length of electromagnetic showers is never parametrically larger than the trigger size.

\paragraph{Neutrinos.}
Neutrinos scatter off nuclei with a cross section that increases with energy.
In these interactions, an $\OO(1)$ fraction of the neutrino energy is transferred to the nucleus with the rest going to produced leptons---this is sufficient to start a hadronic shower \cite{Gandhi:1998ri, Formaggio:2013kya}.
At an energy of $\sim 10^{11} ~\GeV$,~\cite{Gandhi:1998ri} calculates the neutrino-nuclear cross section to be $\sim 10^{-32} ~\cm^2$.
Conservatively assuming this value for even higher energies, we find a neutrino mean free path in a WD of order $\sim 10~\cm$.
Therefore, any high-energy neutrino released in the WD will (on average) only interact after traveling a distance $\gg \lambda_T$. 
As per the discussion above, this makes the heating of a WD via the release of multiple neutrinos highly inefficient due to the (enormous) volume dilution factor in~\eqref{eq:energy_boom_condition}. 
Interestingly, a \emph{single} high-energy neutrino with energy greater than $\Eboom$ will still be able to efficiently heat the star and trigger a runaway. 
This is because the neutrino mean free path is simply a displacement after which a compact shower of size $X_\text{had}$ occurs. 
If the energy contained in a single shower is large enough, then the heating caused by this single neutrino can effectively be considered as a separate and efficient heating event. 

\subsection{Low-Energy Elastic Heating}
The showers of high-energy particles described above terminate in a cloud of low-energy $\epsilon \sim 10~\MeV$ neutrons, protons, and charged pions, and $\epsilon \sim 10-100~\MeV$ electrons and photons.
Of course, particles at these energies may also be directly produced by the DM.
At these energies, elastic nuclear, Coulomb, and Compton scatters dominate and eventually lead to the thermalization of ions.
Once again, the physical expressions for all computed stopping powers and stopping lengths are given in Appendix~\ref{sec:wdpdg} whereas we simply quote the relevant numerical values here. 

\paragraph{Hadrons.}
Neutral hadrons are the simplest species we consider, interacting at low-energies only through elastic nuclear scatters with cross section $\sigma_\text{el} \approx 1~\text{b}$, where $1~\text{b} = 10^{-24}~\cm^2$. 
Note that the large ion mass requires $\sim 10 - 100$ hard scatters to transfer the hadron's energy in the form of a random-walk.
This elastic heating range is
\begin{align}
 \lambda_\text{el} &\approx
 10^{-7} ~\text{cm} \l\frac{10^{32}~\text{cm}^{-3}}{n_\text{ion}}\r,
\end{align}
and is always less than the trigger size.

Charged hadrons are also subject to Coulomb interactions, which would provide the dominant stopping in terrestrial detectors.
In this case, however, Coulomb scatters off degenerate WD electrons are strongly suppressed and charged hadrons predominantly undergo elastic nuclear scatters like their neutral brethren.
This suppression is due to (1) motion of the electrons, which fixes the relative velocity to be $\OO(1)$ and removes the enhancement of Coulomb stopping usually seen at low velocity, and (2) Pauli blocking, which forces the incident particle to scatter only electrons near the top of the Fermi sea.
For an incident particle with velocity $v_\x{in} \ll 1$, the first effect suppresses the stopping power by a factor of $v_\x{in}^2$ relative to that off stationary, non-degenerate electrons and the second by an additional factor of $v_\x{in}$.
Note that there is a small range of energies in which Coulomb scatters off ions dominate the stopping of charged hadrons---either way, both length scales are well below the trigger size.

\paragraph{Electrons and Photons.}
For electrons and photons below $\sim 100 ~\MeV$ the dominant interactions are Coulomb scatters off WD electrons and Compton scatters, respectively.
The length scale of these processes is smaller than any interaction with ions, and so these electrons and photons will thermalize into a compact electromagnetic ``gas" with a size set by the radiative length scale $X_\text{EM}$.
The EM gas will cool and diffuse to larger length scales, eventually allowing thermalization with nuclei via the subdominant Coulomb scatters of electrons off ions.
The photons of the EM gas will not undergo photonuclear showers here, as the gas will cool below $\sim 10~\MeV$ by the time it diffuses out to a size $\lambda_{\gamma A}$.
This gas temperature is initially at most $\sim 100~\MeV$.
At these temperatures the heat capacity is dominated by photons, so as the gas diffuses to a size $\lambda_{\gamma A}$ it cools by a factor $(X_\text{EM}/\lambda_{\gamma A})^{3/4} \sim 10^{-2} - 10^{-1}$.
Note that for temperatures $T$ less than $E_F$, the electrons are partially degenerate and heating proceeds via the thermal tail with kinetic energies $\epsilon \sim E_F + T$.
Therefore, the relevant thermalization process is Coulomb scattering of electrons off ions.

Like the hadronic elastic scatters, an electron Coulomb scattering off ions will occasionally hard scatter, and thus deposit its energy along a random walk. This reduces the stopping length at low energies, yielding
\begin{equation}
\lambda_\text{coul} \approx 10^{-6}~\cm \l \frac{\epsilon}{10 ~\MeV} \r^{3/2} \l \frac{10^{32} ~\cm^{-3}}{n_\text{ion}}\r
\end{equation}
which is below the trigger size.
\section{Dark Matter-Induced Ignition}
\label{sec:dmignition}
Any DM interaction that produces SM particles in a WD has the potential to ignite the star, provided that sufficient SM energy is produced.
The distribution in space, momentum, and species of these SM products is dependent on unknown DM physics and is needed to determine the rate of DM-induced ignition.
This can be done precisely for a specific DM model, as we do for Q-balls in Section~\ref{sec:qballs}.
In this Section, however, we study some general features of DM-WD encounters involving DM that possesses interactions with itself and the SM.
We collect below the basic formulas relating DM model parameters to ignition criteria, SN rate, etc.

DM can generically heat a WD through three basic processes: DM-SM scattering, DM-DM collisions, and DM decays.
For ultra-heavy DM, these processes can be complicated events involving many (possibly dark) final states, analogous to the interactions of heavy nuclei.
In the case of DM-SM scattering, we consider both elastic and inelastic DM scatters off WD constituents, e.g.~carbon ions.
We classify DM candidates into three types according to the interaction that provides the dominant source of heating, and refer to these as scattering, collision, and decay candidates.
We also make the simplifying assumption that the above events are ``point-like", producing SM products in a localized region (smaller than the heating length) near the interaction vertex.
Where this is not the case (as in our elastic scattering and Q-ball constraints, see Sections~\ref{sec:TransitConstraints} and~\ref{sec:qballs}), then the same formalism applies but with the event size added to the stopping length.

The SN rate may be greatly enhanced if DM is captured in the star, so we also consider separately ``transiting DM" and ``captured DM".
In general, there is some loss of DM kinetic energy in the WD.
In the transit scenario, this energy loss is negligible and the DM simply passes through the star.
In the capture scenario, the energy loss is not directly capable of ignition but is sufficient to stop the DM and cause it to accumulate in the star.
Energy loss may be due to a variety of processes, but for simplicity we will focus on an DM-nuclei elastic scattering.
Of course, due to the velocity spread of DM in the rest frame of a WD, there will necessarily be both transiting and captured DM populations in the star.

\subsection{DM Transit}

\paragraph{DM-SM Scattering.}
Runaway fusion only occurs in the degenerate WD interior where thermal expansion is suppressed as a cooling mechanism.
The outer layers of the WD, however, are composed of a non-degenerate gas and it is therefore essential that a DM candidate penetrate this layer in order to ignite a SN.
We parameterize this by a DM stopping power $(dE/dx)_\text{SP}$, the kinetic energy lost by the DM per distance traveled in the non-degenerate layer, and demand that
\begin{align}
\label{eq:CrustCondition}
  \left( \frac{d E}{d x} \right)_\text{SP} \ll
  \frac{m_\chi v^2_\text{esc}}{R_\text{env}},
\end{align}
where $R_\text{env}$ is the nominal size of the non-degenerate WD envelope and $v_\text{esc} \sim 10^{-2}$ is the escape velocity of the WD, at which the DM typically transits the star. 

DM-SM scattering will result in a continuous energy deposit along the DM trajectory (if the interaction is rare enough for this not to be true, then the encounter is analogous to the case of DM decay).
This is best described by a linear energy transfer $(dE/dx)_\text{LET}$, the kinetic energy of SM particles produced per distance traveled by the DM.
If these products have a heating length $L_0$ then the energy deposit must at minimum be taken as the energy transferred along a distance $L_0$ of the DM trajectory.
Importantly, as per the ignition condition~\eqref{eq:energy_boom_condition}, such a deposition is \emph{less} explosive unless $L_0$ is smaller than the trigger size $\lambda_T$.
We thus consider the energy deposited over the larger of these two length scales.
Assuming the energy of the DM is roughly constant during this heating event, the ignition condition is:
\begin{align}
\label{eq:transitexplosion}
  \left( \frac{d E}{d x} \right)_\text{LET} \gtrsim
  \frac{\Eboom}{\lambda_T} \cdot \text{max}
  \left\{\frac{L_0}{\lambda_T}, 1 \right\}^2.
\end{align}
Note that the DM stopping power $(dE/dx)_\text{SP}$ and the linear energy transfer $(dE/dx)_\text{LET}$ are related in the case of elastic scatters, but in general the two quantities may be controlled by different physics.
In addition, a transit event satisfying condition~\eqref{eq:CrustCondition} will have negligible energy loss over the parametrically smaller distances $\lambda_T$ or $L_0$, validating~\eqref{eq:transitexplosion}.

The above condition sums the individual energy deposits along the DM trajectory as though they are all deposited simultaneously.
This is valid if the DM moves sufficiently quickly so that this energy does not diffuse out of the region of interest before the DM has traversed the region.
We therefore require that the diffusion time $\tau_\text{diff}$ across a heated region of size $L$ at temperature $T_f$ be larger than the DM crossing-time:
\begin{align}
  \tau_\text{diff} \sim \frac{L^2}{\alpha(T_f)} \gg
  \frac{L}{v_\text{esc}},
\label{eq:SlowDiffusion}
\end{align}
where $\alpha(T)$ is the temperature-dependent diffusivity. 
This condition is more stringent for smaller regions, so we focus on the smallest region of interest, $L = \lambda_T$.
Then~\eqref{eq:SlowDiffusion} is equivalent to demanding that the escape speed is greater than the conductive speed of the fusion wave front, $v_\text{cond} \sim \alpha(T_f) / \lambda_T$.
Numerical calculations of $v_\text{cond}$ are tabulated in~\cite{Woosley}, and indeed condition~\eqref{eq:SlowDiffusion} is satisfied for all WD densities.

The rate of transit events is directly given by the flux of DM through a WD
\begin{align}
  \Gamma_\text{trans} \sim
  \frac{\rho_{\chi}}{m_\chi} R_\text{WD}^2
  \l\frac{v_\text{esc}}{v_\text{halo}}\r^2 v_\text{halo},
\label{eq:TransitRate}
\end{align}
where $\rho_\chi$ is the DM density in the region of the WD, and $R_\text{WD}$ is the WD radius.
Here $v_\text{halo} \sim 10^{-3}$ is the virial velocity of our galactic halo.
Note the $(v_\text{esc}/v_\text{halo})^2 \sim 100$ enhancement due to gravitational focusing.

We will not consider here captured DM that heats the star via scattering events, as such heating will typically cause ignition before capture occurs.
However, it is possible to cause ignition after capture if the collection of DM leads to an enhanced scattering process.

\paragraph{DM-DM Collisions and DM Decays.}

For a point-like DM-DM collision or DM decay event releasing particles of heating length $L_0$, ignition will occur if the total energy in SM products satisfies condition~\eqref{eq:energy_boom_condition}.
Such an event will likely result in both SM and dark sector products, so we parameterize the resulting energy in SM particles as a fraction $f_\text{SM}$ of the DM mass.
For non-relativistic DM, the DM mass is the dominant source of energy and therefore $f_\text{SM} \lesssim 1$ regardless of the interaction details.
A single DM-DM collision or DM decay has an ignition condition:
\begin{equation}
\label{eq:coldecay}
  m_\chi f_\text{SM} \gtrsim \Eboom \cdot \text{max} \left \{\frac{L_0}{\lambda_T}, 1 \right \}^3.
\end{equation}
Thus the WD is sensitive to annihilations/decays of DM masses $m_\chi \gtrsim 10^{16} ~\GeV$.

DM that is not captured traverses the WD in a free-fall time $t_\text{ff} \sim R_\text{WD}/v_\text{esc}$, and the rate of DM-DM collisions within the WD parameterized by cross section $\sigma_{\chi \chi}$ is:
\begin{align}
  \Gamma^\text{ann}_\text{SN}
  \sim \l \frac{\rho_\chi}{m_\chi} \r^2 \sigma_{\chi \chi} \l \frac{v_\text{esc}}{v_\text{halo}}\r^3 v_\text{halo} R_\text{WD}^3.
  \label{eq:collisionDM}
\end{align}
Similarly the net DM decay rate inside the WD parameterized by a lifetime $\tau_\chi$ is:
\begin{align}
 \Gamma^\text{decay}_\text{SN}
   \sim \frac{1}{\tau_\chi} \frac{\rho_{\chi}}{m_\chi} \l \frac{v_\text{esc}}{v_\text{halo}}\r R_\text{WD}^3.
  \label{eq:decayDM}
\end{align}

\subsection{DM Capture}

\paragraph{Review of DM Capture.}

We first summarize the capture and subsequent evolution of DM in the WD, ignoring annihilations or decays---see Appendix~\ref{sec:capture} for details.
Consider a spin-independent, elastic scattering off carbon ions with cross section $\sigma_{\chi A}$.
The rate of DM capture in gravitating bodies is of course very well-studied~\cite{Press:1985ug, Gould:1987ir}.
However, this rate must be modified when the DM requires multiple scatters to lose the necessary energy for capture.
Ultimately, for ultra-heavy DM the capture rate is of the form
\begin{align}
  \Gamma_\text{cap} &\sim \Gamma_\text{trans} \cdot
  \text{min}\left\{1, \overbar{N}_\text{scat} \frac{m_\text{ion} v_\text{esc}^2}{m_\chi v_\text{halo}^2} \right\},
\end{align}
where $\overbar{N}_\text{scat} \sim n_\x{ion} \sigma_{\chi A} R_\x{WD}$ is the average number of DM-carbon scatters during one DM transit.
For the remainder of this Section, all results are given numerically assuming a WD central density $n_\text{ion} \sim 10^{31}~\cm^{-3}$.
The relevant parametric expressions are presented in further detail in Appendix~\ref{sec:capture}. 

Once DM is captured, it eventually thermalizes with the stellar medium at velocity $v_\text{th} \sim \l T_\x{WD}/m_\chi \r^{1/2}$, where $T_\text{WD}$ is the WD temperature. 
The dynamics of this process depend on the strength of the DM-carbon interaction, namely on whether energy loss to carbon ions provides a small perturbation to the DM's gravitational orbit within the star or whether DM primarily undergoes Brownian motion in the star due to collisions with carbon.
For simplicity, we will focus here only on the former case, corresponding roughly to interactions
\begin{align}
\label{eq:slowcapture}
    \sigma_{\chi A} \lesssim \frac{m_\chi}{\rho_\x{WD} R_\x{WD}}
    \sim 10^{-26}~\x{cm}^2 \; \l \frac{m_\chi}{10^{16}~\GeV} \r
\end{align}
where the DM is able to make more than a single transit through the star before thermalizing.
Note that the opposite regime indeed also provides constraints on captured DM and is unconstrained by other observations, see Figure~\ref{fig:elastic-capture}, however the resulting limits are similar to those presented here.

In the limit~\eqref{eq:slowcapture}, captured DM will thermalize by settling to a radius $R_\x{th}$ given by the balance of gravity and the thermal energy $T_\x{WD}$,
\begin{align}
  R_\text{th} \approx 0.1 ~\cm \l \frac{m_\chi}{10^{16} ~\GeV}\r^{-1/2}.
\end{align}
This settling proceeds in two stages.
Captured DM will initially be found on a large, bound orbit that exceeds the size of the WD, decaying after many transits of the star until the orbital size is fully contained within the WD.
This occurs after a time
\begin{equation}
\label{eq:thermalization1}
t_1 \approx 7\times 10^{16}~\text{s}
  \l \frac{m_\chi}{10^{16} ~\GeV} \r^{3/2}
  \l \frac{\sigma_{\chi A}}{10^{-35} ~\cm^2} \r^{-3/2}.
\end{equation}
The DM then completes many orbits within the star until its orbital size decays to the thermal radius, occurring after a further time
\begin{equation}
\label{eq:thermalization2}
t_2  \approx 10^{14}~\text{s}\l \frac{m_\chi}{10^{16} ~\GeV} \r
  \l \frac{\sigma_{\chi A}}{10^{-35} ~\cm^2} \r^{-1}.
\end{equation}
Note that the difference in scalings between $t_1$ and $t_2$ is due to the fact that, while the two times are ultimately determined by scattering in the star, the dynamics of the settling DM are quite distinct in each case.
$t_1$ is dominated by the time spent on the largest orbit outside the WD (which additionally depends on $\sigma_{\chi A}$) while $t_2$ is dominated by the time spent near the thermal radius.
Subsequently the DM will begin steadily accumulating at $R_\text{th}$, with the possibility of self-gravitational collapse if the collected mass of DM exceeds the WD mass within this volume.
This occurs after a time
\begin{align}
\label{eq:tsg}
t_\text{sg} &\approx
  10^{9} ~\text{s} \l \frac{m_\chi}{10^{16} ~\GeV} \r^{-1/2}
  \l \frac{\sigma_{\chi A}}{10^{-35} ~\cm^2} \r^{-1}.
\end{align}
Of course, not all of these stages may be reached within the age of the WD $\tau_\text{WD}$. 
The full time to collect and begin self-gravitating is $t_1 + t_2 + t_\x{sg}$.

At any point during the above evolution, captured DM has the potential to trigger a SN.
We will consider ignition via either the decay or annihilation of captured DM.
Of particular interest are events occurring within a collapsing DM core, as such cores have the additional ability to ignite a WD for DM masses less than $\Eboom$, either via multiple DM annihilations or by the formation of a black hole.
This is the focus of forthcoming work~\cite{us}.
In the following, we restrict attention to the limit~\eqref{eq:slowcapture} and require DM masses sufficiently large so that a single collision or decay will ignite the star, and give only a quick assessment of DM core collapse.

\paragraph{Captured DM-DM Collisions.}
We now turn to the rate of DM-DM collisions for captured DM.
Of course, the thermalizing DM constitutes a number density of DM throughout the WD volume.
Assuming that $t_1 + t_2 < \tau_\text{WD}$, the total rate of annihilations for this ``in-falling" DM is peaked near the thermal radius and is of order:
\begin{equation}
\label{eq:infall}
\Gamma_\text{infall} \sim \frac{(\Gamma_\text{cap} t_2)^2}{R_\text{th}^3} \sigma_{\chi \chi} v_\text{th}.
\end{equation}
If $\Gamma_\text{infall} t_2 > 1$, then a SN will be triggered by the in-falling DM population.
Otherwise if $\Gamma_\text{infall} t_2 < 1$, the DM will start accumulating at the thermal radius.
If $t_\text{sg} \ll t_2$ (as expected for such heavy DM masses) there will be no collisions during this time and thus a collapse will proceed.
For a DM sphere consisting of $N$ particles at a radius $r$, the rate of annihilations is
\begin{align}
\label{eq:collapse}
\Gamma_\text{collapse} &\sim \frac{N^2}{r^3} \sigma_{\chi \chi} v_\chi, \\
 v_\chi &\sim \sqrt{\frac{G N m_\chi}{r}}.
\end{align}
Of course, there may be some stabilizing physics which prevents the DM from collapsing and annihilating below a certain radius, such as formation of a black hole or bound states.
To illustrate the stringent nature of the collapse constraint we will simply assume some benchmark stable radius, as in Figure~\ref{fig:capture-collision}.
We assume that the timescale for collapse at this radius is set by DM cooling $t_\x{cool}$, which is related to $t_2$.
Note that if a single collision has not occurred during collapse, one may additionally examine annihilations of the subsequent in-falling DM down to the stable radius---for simplicity, we do not consider this scenario.

\paragraph{Captured DM Decays.}
Lastly, we compute the rate of decays for captured DM, which is simply proportional to the number of DM particles in the WD available for decay at any given instance.
In the transit scenario~\eqref{eq:decayDM}, this rate is $\Gamma \sim \tau_\chi^{-1} \Gamma_\text{trans} t_\text{ff}$.
In the capture scenario, this number is instead determined by the thermalization time within the WD $\Gamma \sim \tau_\chi^{-1} \Gamma_\text{cap} t_2$, conservatively assuming that after a thermalization time, the DM quickly collapses and stabilizes to an ``inert" core incapable of further decay.
If this is not the case, then the captured DM decay rate is given by $\Gamma \sim \tau_\chi^{-1} \Gamma_\text{cap} \tau_\text{WD}$.

\section{Dark Matter Constraints}
\label{sec:constraints}
We now constrain some generic DM candidates which will ignite a WD via one of the processes parameterized in Section~\ref{sec:dmignition}.
These release SM particles that deposit their energy and thermalize ions within a distance described in Section~\ref{sec:smheating}.
First, however, we review how WD observables constrain DM candidates capable of triggering SN.

\subsection{Review of WD Observables}
Following the discussion of~\cite{Graham:2015apa}, our constraints come from (1)~the existence of heavy, long-lived white dwarfs, or (2)~the measured type Ia SN rate.
The ages of WD can be estimated by measuring their temperature and modeling their cooling over time. 
This has been extensively studied, see for example~\cite{Winget:1987}, and it is found that typical age of an old WD is of order $\sim 1 \, \text{Gyr}$.
RX~J0648.04418 is one such nearby star and one of the heavier known WDs, with a mass $\sim 1.25 ~M_{\astrosun}$~\cite{Mereghetti:2013nba} and local dark matter density which we take to be $\rho_\chi \sim 0.4 ~\GeV/\text{cm}^3$.
Of course, this is not the only known heavy WD---the Sloan Digital Sky Survey~\cite{SDSS} has found $20+$ others.
The NuStar collaboration has also recently uncovered evidence for the likely existence of heavy WDs near the galactic center~\cite{NuStar}, where the DM density is assumed to be much greater $\rho_\chi \gtrsim 10^3 ~\text{GeV}/\text{cm}^3$~\cite{Nesti:2013uwa}.
Such heavy candidates are particularly suited for our constraints as the energy deposit necessary to trigger SN~\eqref{eq:energy_boom_condition} is a decreasing function of WD mass.
However, less dense white dwarfs are significantly more abundant in the galaxy.
Thus, even if a sufficiently massive DM is unable to trigger a violent heating event within the lifetime of a WD, it could still ignite enough lighter WDs to affect the measured SN rate of $\sim $ 0.3 per century.
The DM-induced SN rate is estimated using the expected number of white dwarfs per galaxy $\sim 10^{10}$ and their mass distribution~\cite{SDSS}.
Simulations indicate that only WD masses heavier than $\sim 0.85 ~M_{\astrosun}$ will result in optically visible SN~\cite{Graham:2015apa}.
Therefore, most of the stars exploded in this manner will be in the mass range $\sim 0.85 - 1 ~M_{\astrosun}$, resulting in weaker SN than expected of typical Chandrasekhar mass WDs.

To summarize, a bound on DM parameters can be placed if either a single explosive event occurs during the lifetime of an observed star such as RX~J0648.04418, or the SN rate due to such DM events throughout the galaxy exceeds the measured value.
Note that for low-mass WDs dominated by photon diffusion, $\Eboom$ is a strong function of WD density.
The average density for WDs is typically a factor $\sim 10^{-2} - 10^{-1}$ less than the central density, although it is found that the WD density only changes by an $\OO(1)$ fraction from the central value up to a distance $\sim R_\text{WD}/2$~\cite{Chandrasekhar}.
Therefore the central density is a valid approximation as long as we consider heating events within this ``modified" WD volume.
For simplicity, we employ this approach.

\subsection{Scattering Constraints}
\label{sec:TransitConstraints}

In order to constrain a DM model with a scattering interaction, we require that it satisfy the ignition condition~\eqref{eq:transitexplosion}.
This is given in terms of an LET, which parameterizes the ability for DM to release sufficient energy to the star in the form of SM particles.
Here we consider a DM elastic scattering off carbon ions with cross section $\sigma_{\chi A}$, which has an LET:
\begin{align}
\label{eq:schematicLET}
  \left( \frac{d E}{d x} \right)_\text{LET} \sim n_\text{ion} \sigma_{\chi A} m_\text{ion} v_\text{esc}^2.
\end{align}
This can be expressed in terms of the cross section per nucleon $\sigma_{\chi n}$---see Appendix \ref{sec:capture}
Each elastic scatter transfers an energy of order $m_\text{ion} v_\text{esc}^2 \approx 1-10~\text{MeV}$ to the target nuclei, thus enabling fusion reactions.
Note that the stopping power of the DM in the non-degenerate envelope is of the same form, but with the density replaced by its diminished value in this region.
It is interesting that combining the ignition condition~\eqref{eq:transitexplosion} with the requirement that the DM adequately penetrates the non-degenerate layer~\eqref{eq:CrustCondition} yields a lower bound on DM mass.
\begin{align}
\label{eq:transitmass}
m_\chi > \Eboom \l \frac{R_\text{env}}{\lambda_T} \r \l \frac{\rho_\text{env}}{\rho_\text{WD}} \r \frac{1}{v_\text{esc}^2},
\end{align}
where $\rho_\text{WD}$ is the central density of the WD.
Here $R_\text{env} \approx 50 ~\text{km}$ is the width of a non-degenerate WD envelope---the density in this region $\rho_\text{env}$ is typically a small fraction $\sim 10^{-3}$ of the central density~\cite{KippenhahnWeigert}.
We conservatively take the envelope to be composed of carbon ions; if it were primarily hydrogen or helium, then the condition for penetration is weakened by 4 orders of magnitude due to the reduced energy transfer and cross section for scattering.
We find that the DM must be heavier than $\sim 10^{28} ~\GeV$ to ensure an explosive transit of a $1.25~M_{\astrosun}$ WD \emph{and} minimal loss of kinetic energy in the non-degenerate layer.
For the sake of comparison this corresponds to a macroscopic DM mass of order $\sim 20~\text{kg}$. 

Of course, this bound is only applicable if the energy input to the WD is solely coming from DM kinetic energy.
We may also consider DM inelastic scattering off carbon ions which transfer more than $\sim \text{MeV}$ per collision.
Examples of such a process include baryon-number violating interactions which can release the nucleon mass energy $\sim \GeV$ per collision.
This is similar to Q-balls, which absorb the baryon number of nuclear targets and liberate binding energy rather than transferring kinetic energy---this interaction is examined in Section~\ref{sec:qballs}.
Note that the assumption of a ``point-like" interaction requires that the physical size of the DM is much smaller than $\lambda_T$---this is sensible up to masses of order $\sim 10^{47}~\GeV$, at which point the gravitational radius of the DM exceeds $\lambda_T$.

\begin{figure}
\includegraphics[scale=.35]{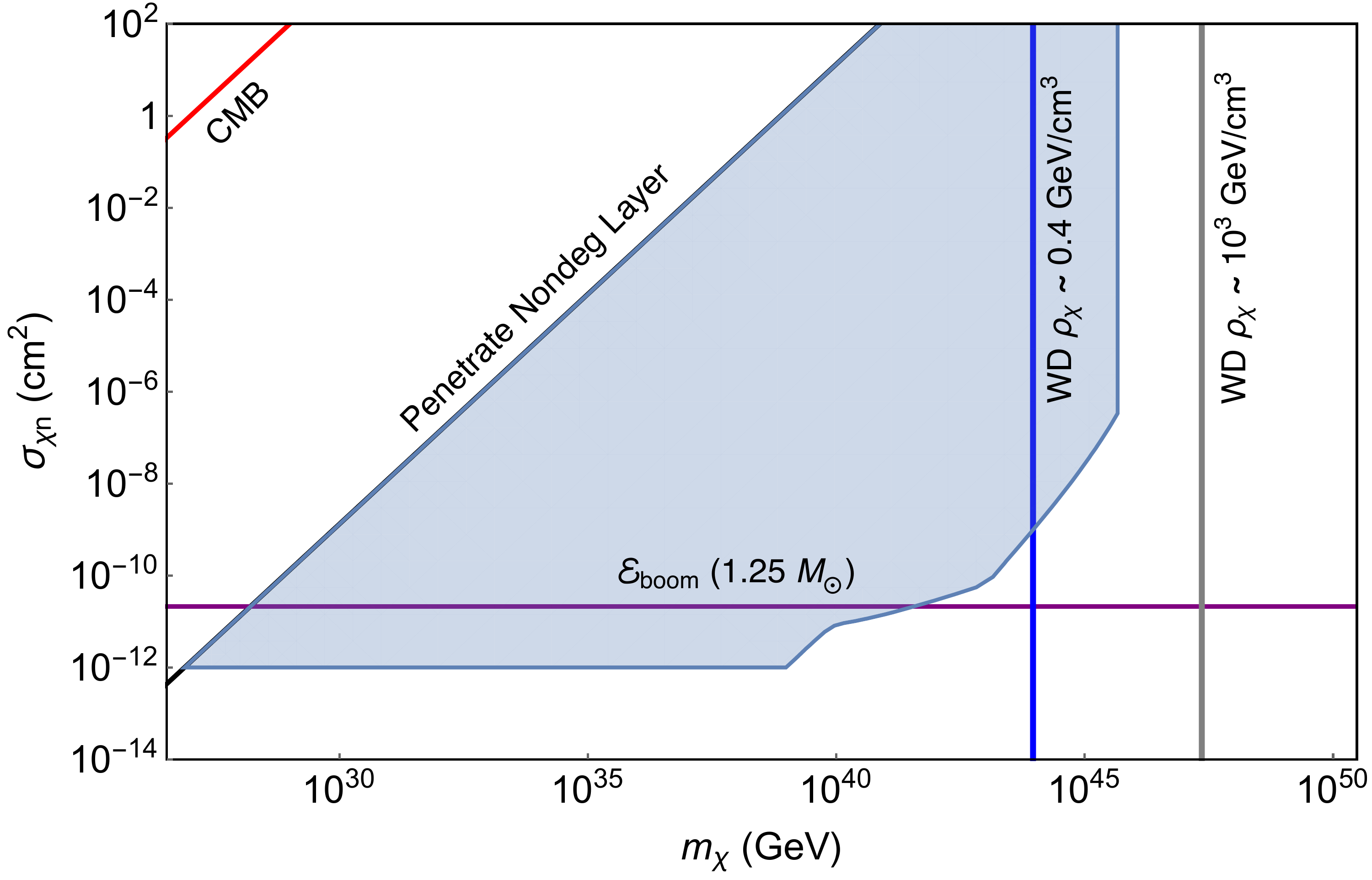}
\caption{Constraints on DM-carbon elastic scattering cross section.
Bounds come from demanding that the DM transit triggers runaway fusion~\eqref{eq:transitexplosion} and occurs at a rate~\eqref{eq:TransitRate} large enough to either ignite a $1.25~M_{\astrosun}$ WD in its lifetime or exceed the measured SN rate in our galaxy (blue shaded).
We also demand that the DM penetrates the non-degenerate stellar envelope, taken at the highest densities, without losing appreciable kinetic energy.
Constraints from the CMB/large-scale structure~\cite{Dvorkin:2013cea} are depicted as well.
}
\label{fig:transit-elastic}
\end{figure}

In Figure~\ref{fig:transit-elastic} we constrain the DM elastic scattering cross section per nucleon $\sigma_{\chi n}$ as a function of DM mass $m_\chi$ using the different classes of observables described above.
Note that the scattering cross sections constrained here are incredibly large $\gtrsim 10^{-10} ~\cm^2$---however, the constraints from WDs reach to very large masses for which no other constraints exist.
At these masses, the most stringent limits on DM elastic scattering are from CMB and Lyman-$\alpha$ spectrum analysis~\cite{Dvorkin:2013cea}, which constrain $\frac{\sigma_{\chi n}}{m_\chi} < \frac{10^{-3} \text{b}}{\GeV}$.
These cross sections also require that the DM involved be macroscopically large, of order or larger than the trigger size, and so the interaction is decidedly not ``point-like.''
This fact does not weaken our constraints, however, since the energy transferred to each ion in the DM's path is greater than $\sim \text{MeV}$.

\subsection{Collision and Decay Constraints}
\label{sec:CollisionConstraints}

\begin{figure}
\includegraphics[scale=.35]{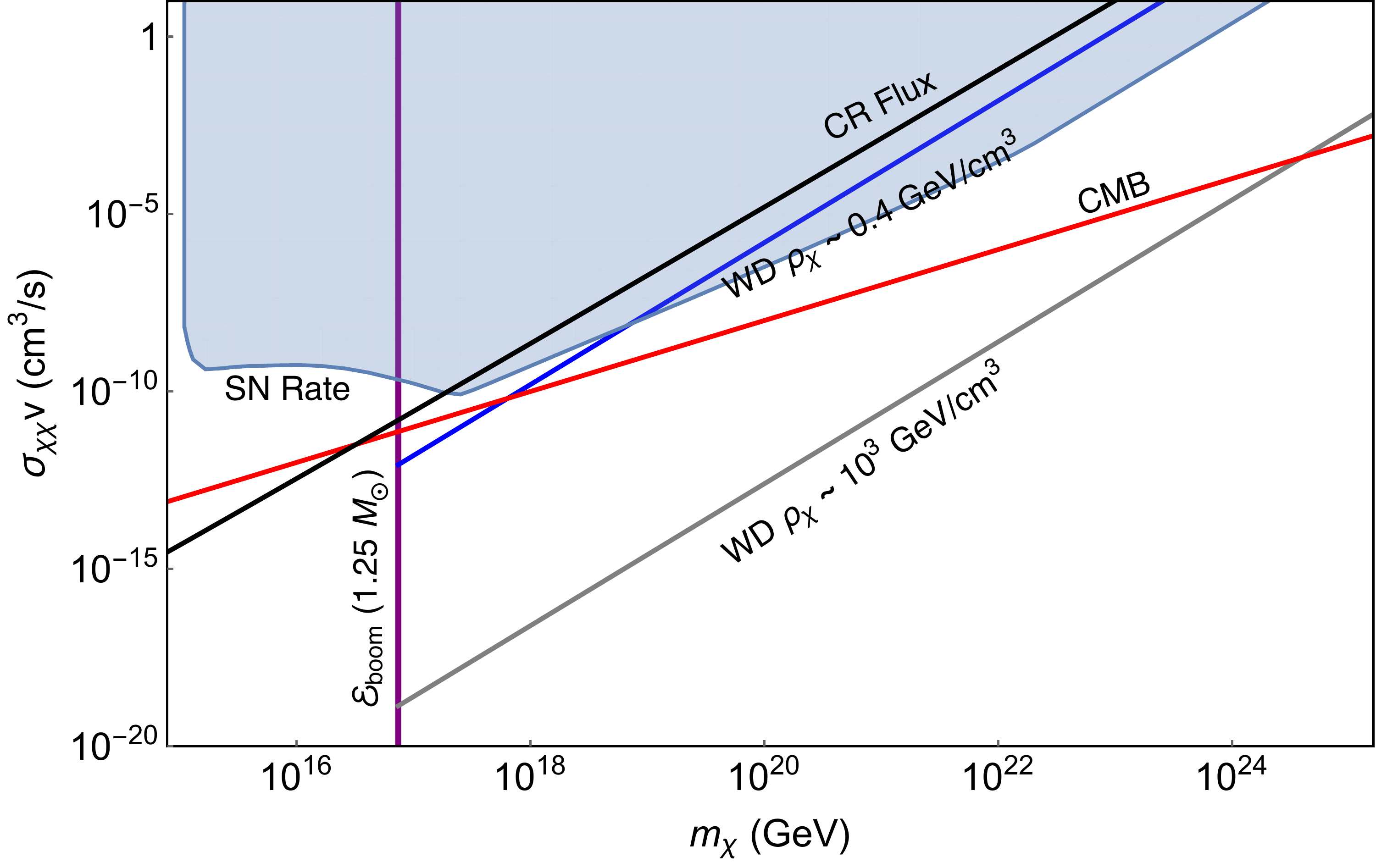}
\caption{Constraints on DM-DM collision cross section to SM products of energy $\epsilon \gg \MeV$.
Bounds come from demanding that the DM transit interaction triggers runaway fusion~\eqref{eq:coldecay} and occurs at a rate~\eqref{eq:collisionDM} large enough to either ignite an observed $1.25~M_{\astrosun}$ WD in its lifetime or exceed the measured SN rate in our galaxy (blue shaded).
Also shown are the CMB \cite{Slatyer:2009yq} (red) and CR flux (black) constraints on DM annihilations.}
\label{fig:transit-collision}
\end{figure}

\begin{figure}
\includegraphics[scale=.35]{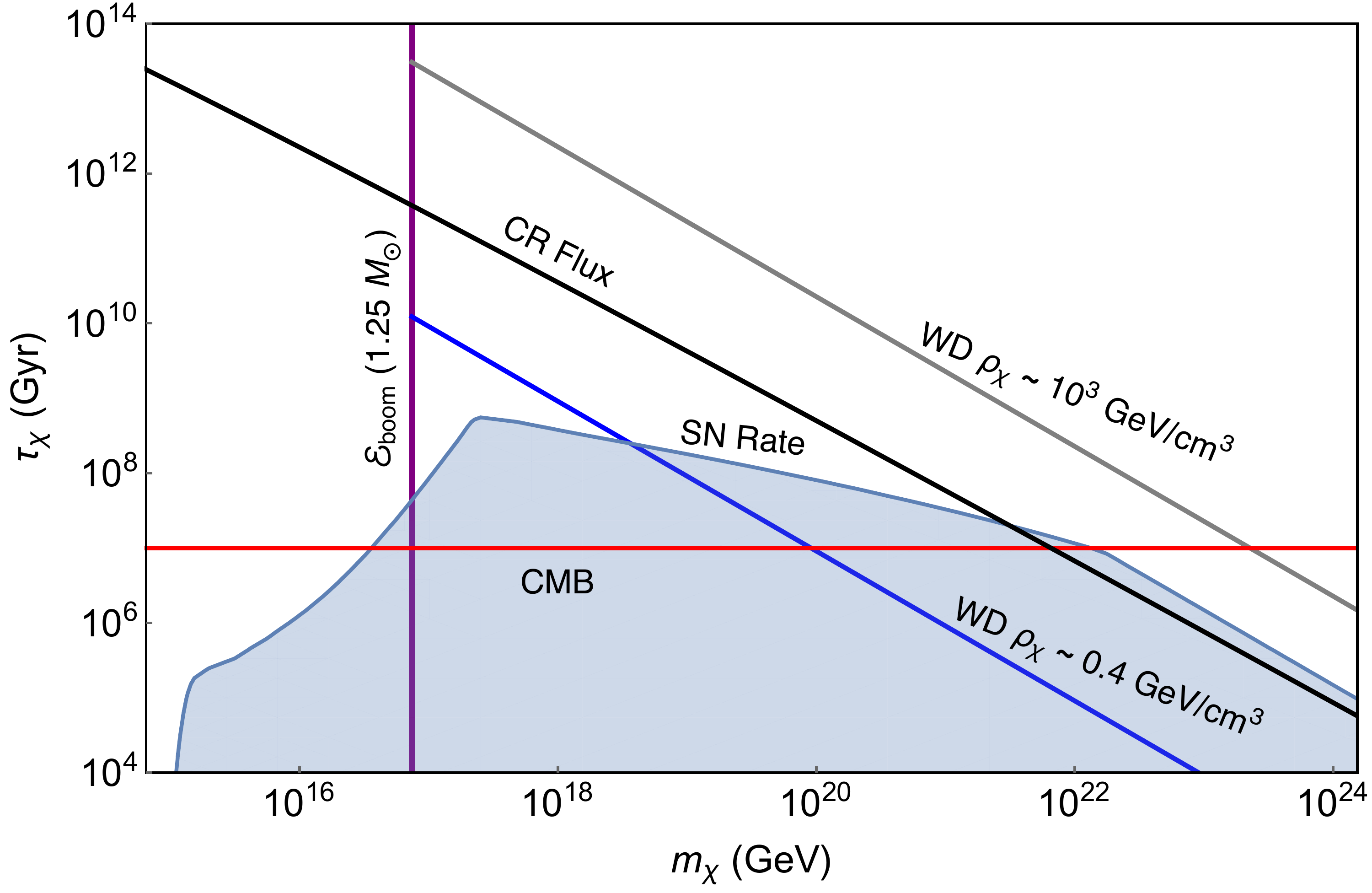}
\caption{Constraints on DM decay to SM products of energy $\epsilon \gg \MeV$.
Bounds come from demanding that the DM transit interaction triggers runaway fusion~\eqref{eq:coldecay} and occurs at a rate~\eqref{eq:decayDM} large enough to either ignite an observed $1.25~M_{\astrosun}$ WD in its lifetime or exceed the measured SN rate in our galaxy (blue shaded).
Also shown are the CMB \cite{Slatyer:2016qyl} (red) and CR flux (black) constraints on DM lifetime.}
\label{fig:transit-decay}
\end{figure}

In order to constrain a DM model through its annihilations or decays within a WD, we require that it satisfy the ignition condition~\eqref{eq:coldecay}.
Consider a single annihilation or decay with $f_\text{SM} = 1$ that releases a spectrum of SM particles.
As shown in Section~\ref{sec:smheating}, the constraint has minimal dependence on the released species if the typical energy $\epsilon$ of secondary products is greater than an MeV.
In the case of neutrinos, we may simply demand that $\epsilon$ is sufficiently large that a single neutrino can ignite the star.
With this schematic for the DM interaction, we can constrain the cross section for collision $\sigma_{\chi \chi}$ and lifetime $\tau_\chi$.
This is done in Figures~\ref{fig:transit-collision} and~\ref{fig:transit-decay} in the case of transiting DM using the different classes of observables for DM-DM collisions and DM decays, respectively.

Of course there are existing limits on DM annihilations and decays, complementary to the ones placed from WDs.
DM annihilations/decays inject energy and affect the ionization history of our universe, which can be probed by measurements of the CMB temperature and polarization angular spectrum \cite{Padmanabhan:2005es, Slatyer:2009yq, Slatyer:2016qyl}.
These constraints are of order $\sigma_{\chi \chi} v < 10^{-27} ~\frac{\cm^3}{\text{s}} \l \frac{m_\chi}{10 ~\GeV} \r$ for annihilations, and  $\tau_\chi > 10^{7} ~\text{Gyr}$ for decay.
There are also constraints on DM annihilation/decays in our halo from the cosmic ray (CR) flux seen in large terrestrial detectors.
Here we provide a crude estimate of the expected constraints from CRs in the case of DM annihilation (decays are qualitatively similar).
A more detailed analysis is beyond the scope of this work.
The Pierre Auger Observatory \cite{ThePierreAuger:2015rma} has detected the flux of $E_\text{th} \sim 10^{11} ~\GeV$ cosmic rays with an exposure of order $A_\text{PA} \sim 40000 ~\text{km}^2 ~\text{sr} ~\text{yr}$.
Ultra-heavy DM annihilations $m_\chi > 10^{16} ~\GeV$ will generally produce secondary particles of energy $\epsilon \gtrsim E_\text{th}$ via final-state radiation.
For a simple 2-2 process (e.g. $\chi \chi \to q q$), the expected number of final-state particles radiated at $E_\text{th}$ due to QCD showers is approximated by the Sudakov double logarithm
\begin{equation}
N_\text{rad} \sim \frac{4 \alpha_s}{\pi} \log\l\frac{m_\chi}{\Lambda_\text{QCD}}\r \log\l\frac{m_\chi}{E_\text{th}}\r \approx 100,
\end{equation}
where $\alpha_s$ is the QCD coupling constant.
Similarly, the estimated number of final-state particles at $E_\text{th}$ due to EW showers is $\approx 50$.
We expect that CRs at this energy originating in our galaxy will be able to strike the earth unattenuated.
Thus, such events would affect the measured CR flux of Pierre Auger unless
\begin{equation}
\l \frac{\rho_\chi}{m_\chi}\r^2 \sigma_{\chi \chi} v \frac{R_\text{halo}}{4 \pi} N_\text{rad} \times A_\text{PA} \lesssim 1.
\end{equation}
Here we assume an average value for DM density $\rho_\chi \approx 0.4~\GeV/\cm^3$ as a reasonable approximation to the integral over our galactic halo volume.
Surprisingly, the above CR constraints are (within a few orders of magnitude) comparable to the constraints due to the observation of long-lived WDs.
This is actually due to a coincidence in the effective ``space-time volumes" of the two systems.
A terrestrial CR detector such as Pierre Auger sees events within a space-time volume $(R_\text{det}^2 R_\text{halo} \times t_\text{det})$, where $R_\text{det} \sim 50 ~\text{km}$, $R_\text{halo} \sim 10 ~\text{kpc}$, and $t_\text{det} \sim 10 ~\text{yr}$.
This is similar in magnitude to the WD space-time volume $(R_\text{WD}^3 \times \tau_\text{WD})$.

\begin{figure}
\includegraphics[scale=.35]{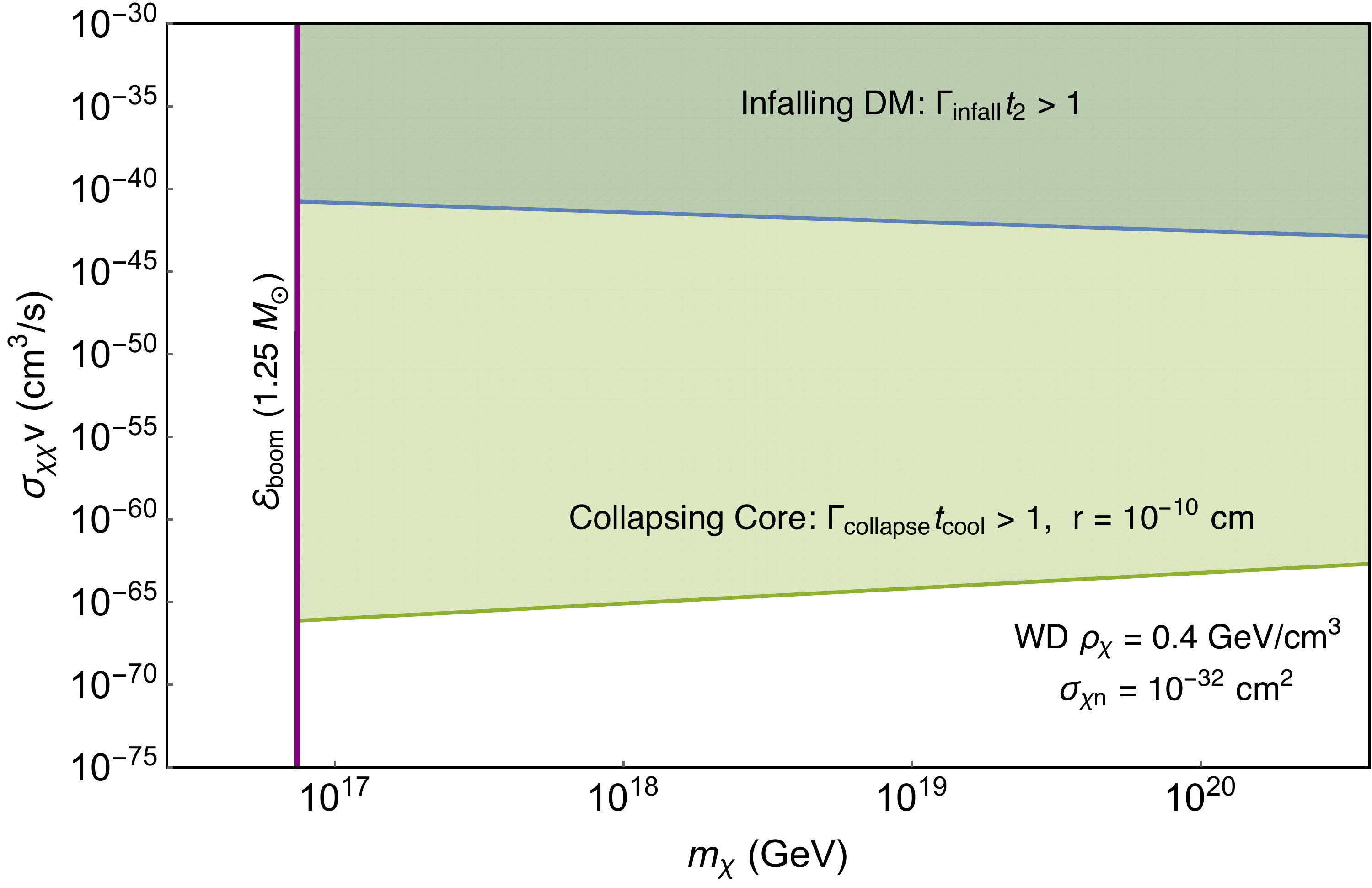}
\caption{Constraints on DM-DM collision cross section to SM products of energy $\epsilon \gg \MeV$, assuming DM is captured with an elastic scattering cross section $\sigma_{\chi n} = 10^{-32} ~\cm^2$.
Bounds come from the observation of $1.25~M_{\astrosun}$ WDs in local DM density.
We consider the annihilation rate during the in-falling thermalization stage~\eqref{eq:infall} (blue shaded) and during self-gravitational collapse~\eqref{eq:collapse} to a stable radius $r = 10^{-10} ~\cm$ (green shaded). See text for details.
}
\label{fig:capture-collision}
\end{figure}

\begin{figure}
\includegraphics[scale=.35]{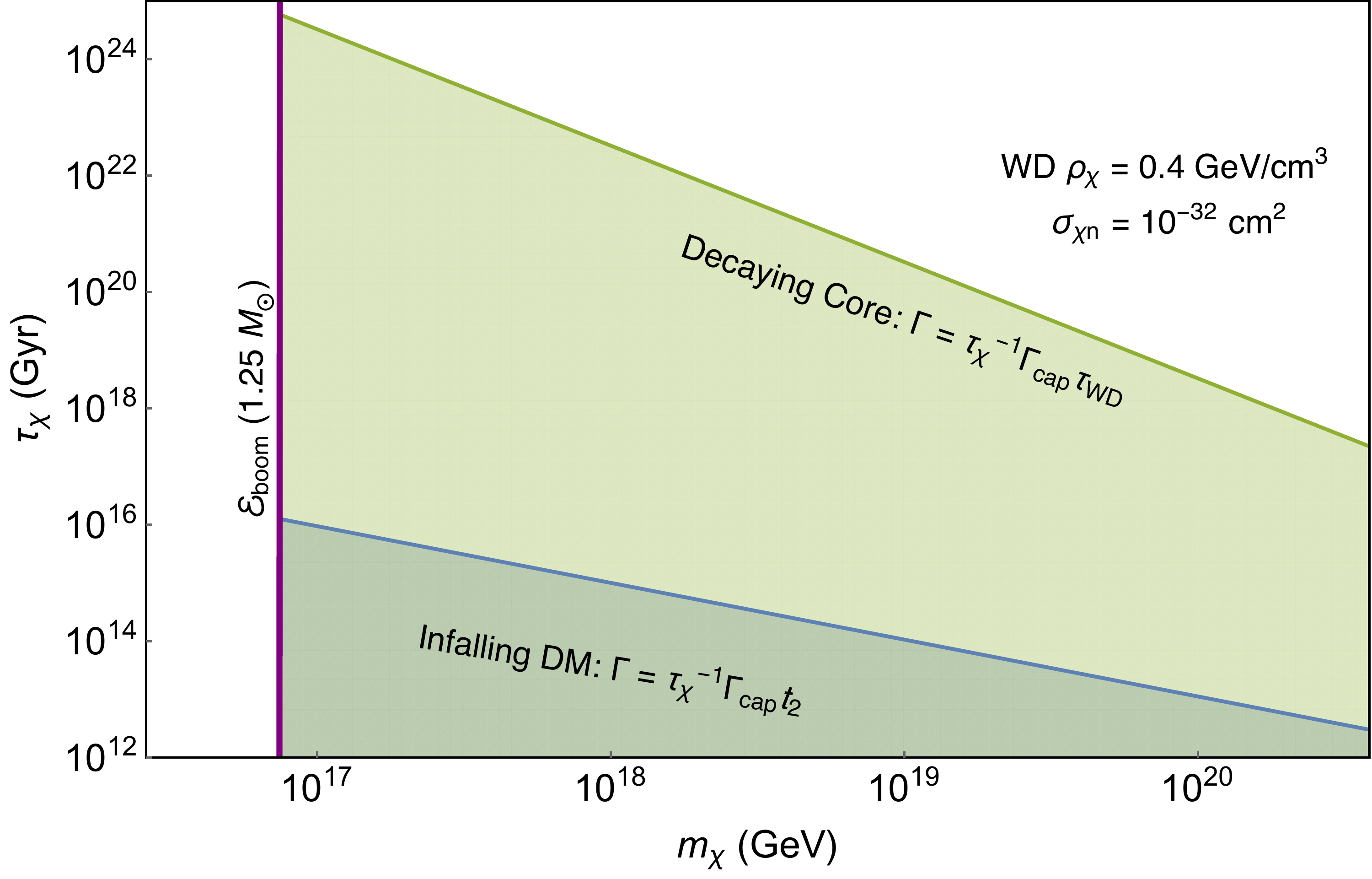}
\caption{Constraints on DM decay to SM products of energy $\epsilon \gg \MeV$, assuming DM is captured with an elastic scattering cross section $\sigma_{\chi n} = 10^{-32} ~\cm^2$.
Bounds come from the observation of $1.25~M_{\astrosun}$ WDs in local DM density.
We consider the rate of decays during the in-falling thermalization stage (blue shaded) and for a decaying DM core (green shaded). See text for details.
}
\label{fig:capture-decay}
\end{figure}

In the case of captured DM, we show the constraints on $\sigma_{\chi \chi}$ and $\tau_\chi$ assuming a benchmark value of the elastic scattering cross section $\sigma_{\chi n} = 10^{-32} ~\cm^2$.
With regards to DM-DM collisions, we also assume a stabilizing radius for the collapsing DM sphere.
This is done in Figures~\ref{fig:capture-collision} and~\ref{fig:capture-decay}---for simplicity, here we only show the constraints from the existence of nearby, heavy WDs.

\begin{figure}
\includegraphics[scale=.35]{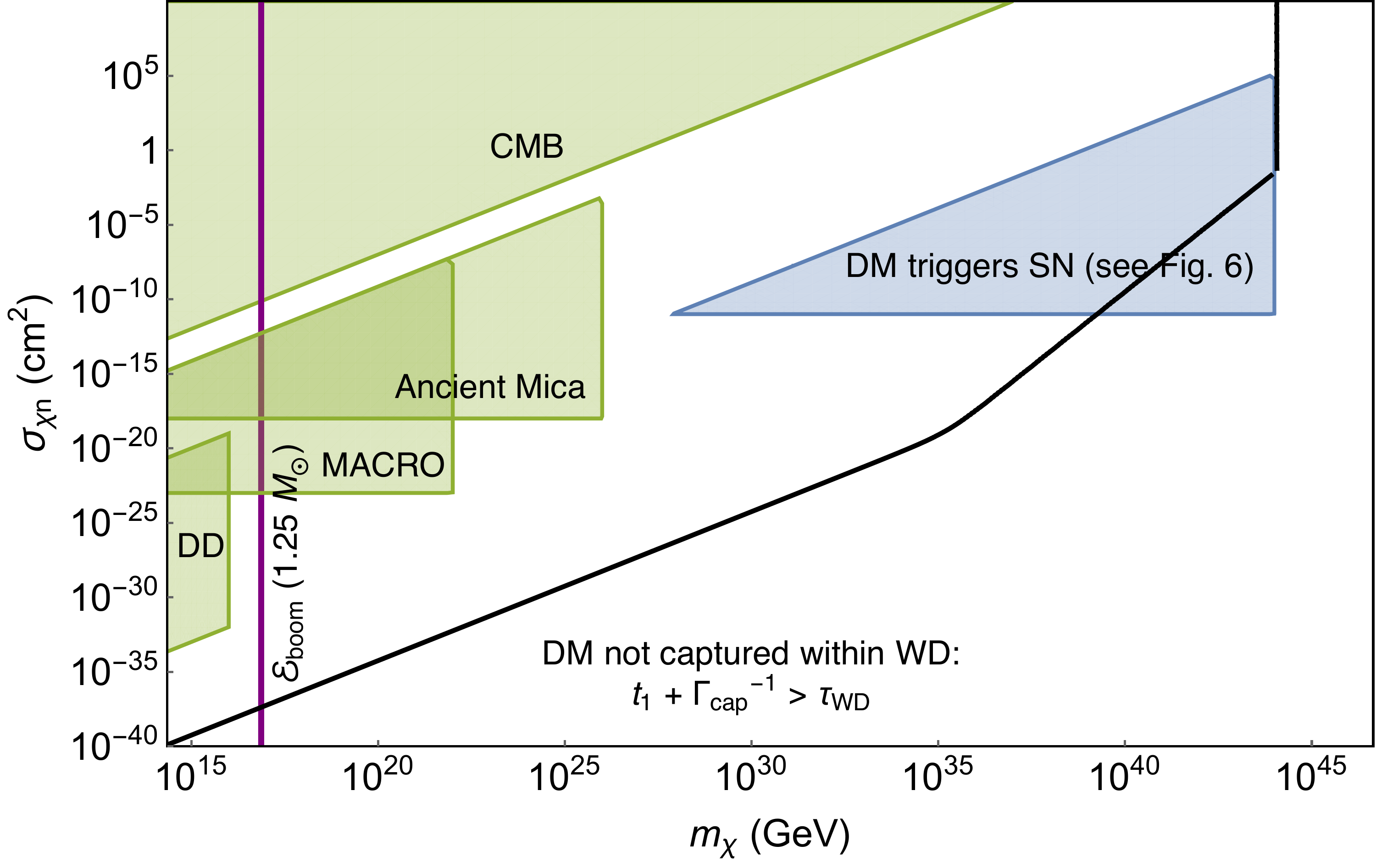}
\caption{Viable parameter space (above the black line) in which DM-nucleon elastic scattering leads to DM capture in a $1.25 ~M_{\astrosun}$ WD.  All of this space is subject to constraints on DM decay and DM-DM annihilation analogous to those given in Figures~\ref{fig:capture-decay} and~\ref{fig:capture-collision}.  Note the blue region, reproducing Figure~\ref{fig:transit-elastic}, indicates DM which causes SN via elastic heating.
We also indicate here estimates of the scattering constraints from cosmology, direct detection, MACRO, and ancient mica \cite{Jacobs:2014yca}.}
\label{fig:elastic-capture}
\end{figure}

It is important to note that there is a large parameter space in $\sigma_{\chi n}$ which will lead to DM capture, thermalization, and core collapse in a WD.
This is depicted in Figure~\ref{fig:elastic-capture}, along with the existing constraints on DM elastic scattering.
As detailed in~\cite{Mack:2007xj}, direct detection experiments such as Xenon 1T~\cite{Aprile:2017iyp} are only sensitive to DM masses $m_\chi < 10^{17} ~\GeV$.
For even larger masses $m_\chi < 10^{26} ~\GeV$ there are constraints from the MACRO experiment \cite{Ambrosio:2002qq} and from ancient excavated mica.
The latter has been studied in~\cite{Jacobs:2014yca}.
We have similarly estimated the bounds from MACRO assuming a detectable threshold of $\sim 5~\text{MeV}/\cm$~\cite{Ambrosio:2002qq}.

\section{Q-balls}
\label{sec:qballs}
Having derived constraints on generic models of ultra-heavy DM, we turn towards a concrete example.
In various supersymmetric extensions of the SM, non-topological solitons called Q-balls can be produced in the early universe~\cite{Coleman:1985ki, Kusenko:1997si}.
If these Q-balls were stable, they would comprise a component of the DM today.
For gauge-mediated models with flat scalar potentials, the Q-ball mass and radius are given by
\begin{equation}
\label{eq:Qballprop}
M_Q \sim m_S Q^{3/4}, ~~~ R_Q \sim m_S^{-1} Q^{1/4},
\end{equation}
where $m_S$ is related to the scale of supersymmetry breaking, and $Q$ is the global charge of the Q-ball---in our case, baryon number.
The condition $M_Q/Q < m_p$ ensures that the Q-ball is stable against decay to nucleons.
The interaction of relic Q-balls with matter depends on its ability to retain electric charge~\cite{Kusenko:1997vp}. 
We restrict our attention to electrically neutral Q-balls, which induce the dissociation of incoming nucleons and in the process absorb their baryonic charge.
During this proton decay-like process, excess energy of order $\Lambda_\text{QCD}$ is released via the emission of 2--3 pions. 
We assume that for each Q-ball inelastic collision, there is equal probability to produce $\pi^0$ and $\pi^\pm$ under the constraint of charge conservation.
The cross section for this interaction is approximately geometric
\begin{align}
\sigma_Q \sim \pi R_Q^2,
\end{align}
and thus grows with increasing $Q$.
Note that a sufficiently massive Q-ball will become a black hole if $R_Q \lesssim G M_Q$.
In the model described above, this translates into a condition $(M_\text{pl}/m_S)^4 \lesssim Q$.

We now determine the explosiveness of a Q-ball transit.
This process is described by a linear energy transfer
\begin{equation}
\label{eq:QballLET}
\l\frac{dE}{dx}\r_\text{LET} \sim n_\text{ion} \sigma_Q N_\pi \epsilon,
\end{equation}
where the nuclear interaction results in $N_\pi \approx 30$ pions released, each with kinetic energy $\epsilon \approx 500 ~\text{MeV}$.
These pions induce hadronic showers which terminate in low-energy hadrons that rapidly transfer their energy to ions via elastic scatters, as discussed in Section~\ref{sec:smheating}.
The pions have a heating length $X_\text{had} \lesssim \lambda_T$; however, we will see the Q-ball has a finite size $R_Q \gtrsim X_\text{had}$ in the region we are able to constrain.
So, as mentioned in Section~\ref{sec:dmignition}, we take the heating length to be $L_0 \sim R_Q + X_\text{had} \sim R_Q$.
The ignition condition is then given by equations~\eqref{eq:transitexplosion} and~\eqref{eq:QballLET}:
\begin{equation}
 R_Q^2 \gtrsim \frac{1}{n_\text{ion}} \frac{\Eboom}{\lambda_T}
 \; \xmax\left \{ \frac{R_Q}{\lambda_T}, 1\right\}^2
 \l \frac{1}{10~\GeV} \r.
\end{equation}
This implies $\sigma_Q \gtrsim 10^{-12} ~\text{cm}^2$ is sufficient to ignite a $1.25 ~M_{\astrosun}$ WD, which corresponds to a charge $Q \gtrsim 10^{42} ~(m_S/\text{TeV})^4$.
Note that for sufficiently large $Q$, the radius will grow larger than $\lambda_T$.
This situation still results in ignition, however, as the energy $\sim 10~\GeV$ released per ion is much larger than the $\sim \MeV$ needed per ion for fusion.
Note finally that the Q-ball interaction described above results in minimal slowing for Q-balls this massive, so transits will easily penetrate the non-degenerate WD envelope~\eqref{eq:CrustCondition}.

\begin{figure}
\includegraphics[scale=.35]{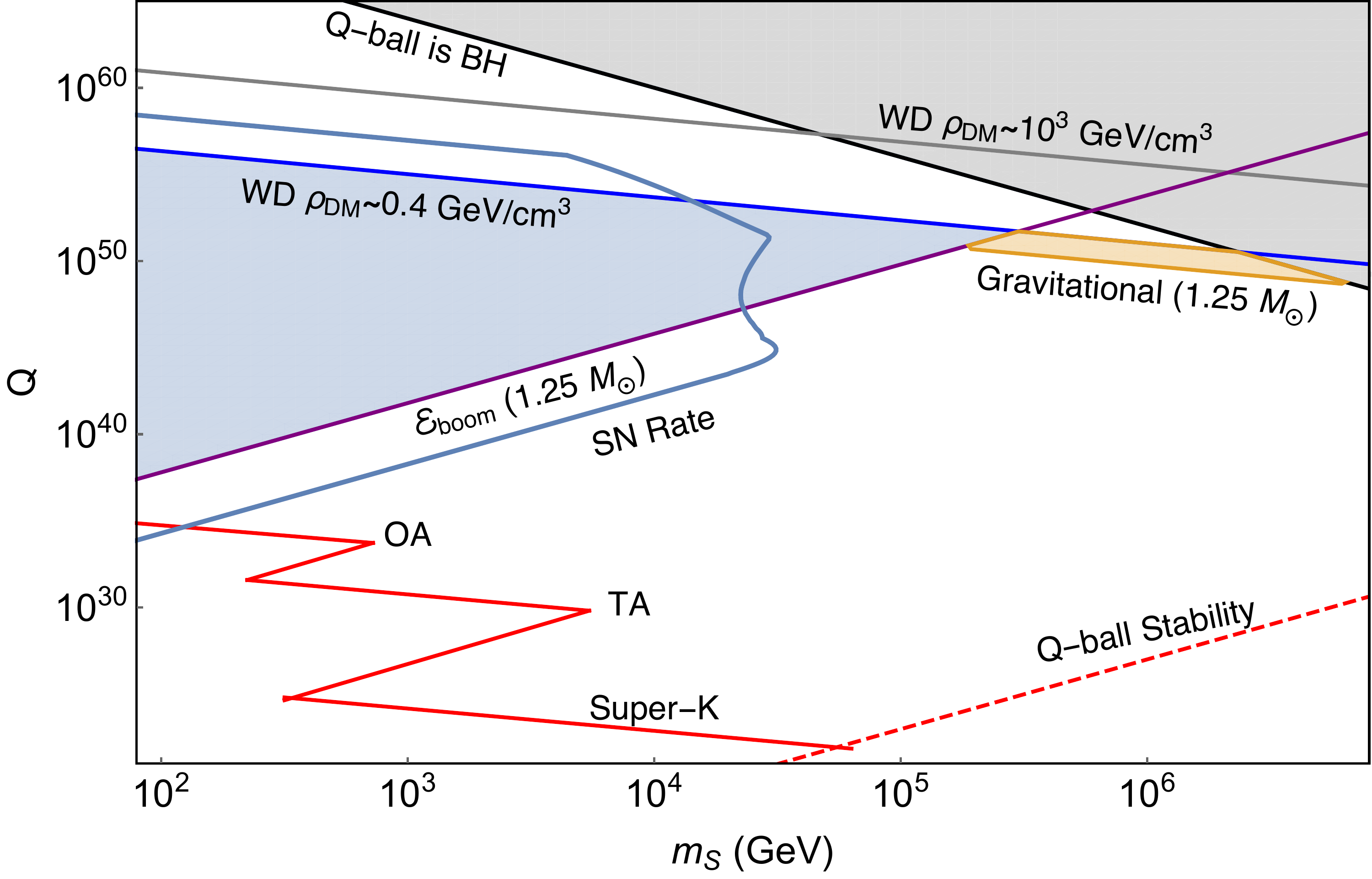}
\caption{
Constraints on Q-ball DM.
Bounds come from demanding that the Q-ball interaction during a DM transit is capable of igniting WDs, occurring at a rate large enough to either ignite a single observed $1.25~M_{\astrosun}$ WD in its lifetime (WD in local DM density is blue shaded) or exceed the measured SN rate in our galaxy.
Also shown is the corresponding constraint from gravitational heating of WDs (orange shaded), and existing limits from terrestrial detectors (red)~\cite{Dine:2003ax}.}
\label{fig:Qballconstraint}
\end{figure}

The existing limits on Q-balls primarily come from Super-Kamiokande and air fluorescence detectors of cosmic rays (OA, TA)~\cite{Dine:2003ax}.
However, the constraints that come from considering the ignition of WDs are in a fundamentally new and complementary region of parameter space.
These are plotted in Figure~\ref{fig:Qballconstraint}.
We have also included the constraints that result from gravitational heating of a WD during a Q-ball transit, as in~\cite{Graham:2015apa}.

\begin{figure}
\includegraphics[scale=1.0]{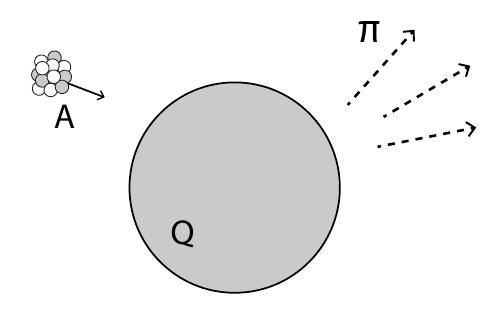}
\caption{Interaction of a baryonic Q-ball with a nucleus $A$. The Q-ball destroys the nucleus and absorbs its baryonic charge, while the excess energy is radiated into roughly $A$ outgoing pions of energy $\Lambda_\text{QCD}$.}
\label{fig:qball-cartoon}
\end{figure}

\section{Discussion}
\label{sec:discussion}
The detection of ultra-heavy DM is an open problem which will likely require a confluence of astrophysical probes.
Here we present a guide to constraining these candidates through DM-SM scatters, DM-DM annihilations, and DM decays inside a WD that release sufficient SM energy to trigger runaway fusion.
In particular, we calculate the energy loss of high-energy particles due to SM interactions within the WD medium and determine the conditions for which a general energy deposition will heat a WD and ignite SN.
Ultra-heavy DM that produces greater than $10^{16}~\GeV$ of SM particles in a WD is highly constrained by the existence of heavy WDs and the measured SN rate.
The formalism provided will enable WDs to be applied as detectors for any DM model capable of heating the star through such interactions.
We have done so for baryonic Q-balls, significantly constraining the allowed parameter space in a complementary way to terrestrial searches.

We have explored briefly the application of this WD instability to self-gravitational collapse of DM cores, which has very interesting possibilities.
The decay or annihilation of DM which is captured by a WD and forms a self-gravitating core is highly constrained for DM with mass greater than $10^{16}~\GeV$.
In addition, such collapsing cores can provide enough heating via multiple annihilations to ignite the star for much smaller DM masses than those considered here, e.g.~$10^7~\GeV$, and can induce SN through other means such as the formation and evaporation of mini black holes.
These will be addressed in future work~\cite{us}.

Finally, in addition to the constraints mentioned above, the general phenomenology of these DM-induced runaways will be the ignition of sub-Chandrasekhar mass WDs, possibly with no companion star present.
Some of the mechanisms considered above are also likely to initiate fusion far from the center of the star.
This is in contrast with conventional single-degenerate and double-degenerate mechanisms, which require a companion star and ignite fusion near the center of a super-Chandrasekhar mass WD~\cite{Maoz:2012}.
This raises the tantalizing possibility that DM encounters with WDs provide an alternative explosion mechanism for type Ia SN or similar transient events, and that these events may be distinguishable from conventional explosions.
Understanding and searching for possible distinguishing features of DM-induced events is an important follow-up work.

\begin{appendices}
\renewcommand{\thesubsection}{\arabic{subsection}}
\section{Particle Stopping in a White Dwarf}
\label{sec:wdpdg}
Here we provide a more detailed analysis of the stopping power (energy loss per distance traveled) of high-energy SM particles in a carbon-oxygen WD due to strong and electromagnetic interactions.
We consider incident electrons, photons, pions, and nucleons with kinetic energy greater than an $\MeV$.

\subsection{WD Medium}
For the WD masses that we consider, the stellar medium consists of electrons and fully-ionized carbon nuclei with central number densities in the range $n_e = Z n_\ion \sim 10^{31} - 10^{33} ~\cm^{-3}$ where $Z=6$.
The internal temperature is $T \sim \keV$~\cite{KippenhahnWeigert}.
The electrons are a degenerate and predominantly relativistic free gas, with Fermi energy
\begin{equation}
  E_F = (3 \pi^2 n_e)^{1/3} \sim 1 -10 ~\MeV.
\end{equation}
The carbon ions, however, are non-degenerate and do not form a free gas.
The plasma frequency due to ion-ion Coulomb interactions is given by
\begin{align}
\Omega_p = \l \frac{4 \pi n_\ion Z^2 \alpha}{m_\ion}\r^{1/2} \sim 1 - 10~\keV,
\end{align}
where $m_\ion$ is the ion mass.
Finally, the medium also contains thermal photons, though these are never significant for stopping particles as the photon number density $n_\gamma \sim T^3$ is much smaller than that of electrons or ions.

\subsection{Nuclear Interactions}
\label{sec:nuclear}

\paragraph{Elastic Scattering of Hadrons.}
Hadrons with energy less than the nuclear binding energy $E_\text{nuc} \sim 10~\MeV$ will predominantly stop due to elastic nuclear scatters with ions.
These are hard scatters, resulting in a stopping power
\begin{align}
  \frac{dE}{dx} \sim n_\ion \sigma_\el
\l \frac{m}{m_\ion}\r E
  \end{align}
for a hadron of mass $m \ll m_\ion$ and kinetic energy $E$.
$\sigma_\el$ is the elastic nuclear scattering cross section, which is of order $\sigma_\el \approx \bn$ at these energies and drops to $\sigma_\el \approx 0.1~\bn$ above $10~\MeV$~\cite{Tavernier}, ignoring the nontrivial effect of nuclear resonances in the intermediate regime $1 - 10~\MeV$.

\paragraph{Inelastic Scattering of Hadrons.}
For energies above $E_\text{nuc}$, the stopping of hadrons is dominated by inelastic nuclear scatters.
In such a collision, an incoming hadron interacts with one or more nucleons to produce a $\OO(1)$ number of additional hadrons which approximately split the initial energy.
At incident energy greater than $\sim \GeV$, the majority of secondary hadrons are pions with transverse momenta $\sim 100 ~\MeV$~\cite{Tavernier}.
Below $\sim \GeV$, it is found that roughly equal fractions of protons, neutrons, and pions are produced in each collision~\cite{Pionnuclear}.
We will thus have a roughly collinear shower terminating at an energy $\sim 10~\MeV$ which consists of pions for most of the shower's development and converts to an mix of pions and nucleons in the final decade of energy.
This cascade is described by a radiative stopping power
\begin{equation}
\label{eq:nucshower}
  \frac{dE}{dx} \sim n_\ion \sigma_\inel E,
\end{equation}
where the inelastic nuclear cross section is given by $\sigma_\inel \approx 100 ~\mbn$ and roughly constant in energy~\cite{Tavernier}.
The total length of the shower is only logarithmically dependent on the initial hadron energy $E$,
\begin{align}
    X_\x{had} \sim \frac{1}{n_\ion \sigma_\inel} \log\l\frac{E}{E_\text{nuc}}\r.
\end{align}

\paragraph{Photonuclear Interactions.}
Photons of energy greater than $10 ~\MeV$ can also strongly interact with nuclei through the production of virtual quark-antiquark pairs.
This is the dominant mode of photon energy loss at high energy.
The photonuclear scatter destroys the photon and fragments the nucleus, producing secondary hadrons in a shower analogous to that described above.
The photonuclear cross section $\sigma_{\gamma A}$ is roughly given by $\sigma_{\gamma A} \approx \alpha \sigma_\inel$, again ignoring the nuclear resonances that occur for $E \lesssim \GeV$~\cite{Tavernier}.
For $E \gtrsim \GeV$, $\sigma_{\gamma A}$ is likely a slowly increasing function of energy due to the coherent interaction of the photon over multiple nucleons~\cite{Gerhardt:2010bj}, however, instead of extrapolating this behavior we conservatively take a constant photonuclear cross section $\sigma_{\gamma A} \approx 1~\mbn$.

\paragraph{Electronuclear Interactions.}
Electrons can similarly lose energy to nuclei by radiating a virtual photon that undergoes a photonuclear scatter, which indeed provides the dominant energy loss for high energy electrons.
The cross section for this process is roughly given by the photonuclear cross section, scaled by a factor representing the probability to radiate such a photon.
This can be estimated with the Weizsacker-Williams approximation, which gives a stopping power that is suppressed from the photonuclear result by $\alpha$ but enhanced by an $\OO(10)$ logarithmic phase space factor~\cite{Gerhardt:2010bj}:
\begin{align}
    \frac{dE}{dx} \sim \alpha \; n_\ion \sigma_{\gamma A} E
    \log\l\frac{E}{m_e}\r.
\end{align}
Unlike the photonuclear interaction, the electronuclear event is a radiative process that preserves the original electron while leaving hadronic showers in its wake.

\subsection{Radiative Processes}
\label{sec:emshowers}

Electromagnetic showers due to successive bremsstrahlung and pair production events off carbon ions are the dominant stopping mechanisms for intermediate-energy electrons and photons.
Both of these processes result in radiative stopping powers, derived semi-classically as~\cite{Klein:1998du}
\begin{equation}
\label{eq:SemiclassicalBrem}
\frac{dE}{dx} \sim \frac{E}{X_0}, ~~~~ X_0^{-1} = 4 n_\ion Z^2 \frac{\alpha^3}{m_e^2} \log{\Lambda}.
\end{equation}
$X_0$ is the well-known radiation length, and $\log\Lambda$ is a Coulomb form factor given by the range of effective impact parameters $b$:
\begin{align}
  \Lambda = \frac{b_\xmax}{b_\xmin}.
\end{align}
The maximal impact parameter is set by the plasma screening length (see~\ref{sec:coulomb_ion}) and the minimum by the electron mass, below which the semi-classical description breaks down.
Note that for the highest WD densities $\Lambda \lesssim 1$, in which case~\eqref{eq:SemiclassicalBrem} ought be replaced by a fully quantum mechanical result as in~\cite{Bethe1934}.
This still results in a radiative stopping power, and so for simplicity we employ~\eqref{eq:SemiclassicalBrem} with $\log{\Lambda} \sim \OO(1)$ for all WD densities.

\paragraph{LPM Suppression}
A radiative event involving momentum transfer $q$ to an ion must, quantum mechanically, occur over a length $\sim q^{-1}$.
All ions within this region contribute to the scattering of the incident particle, and for sufficiently small $q$ this results in a decoherence that suppresses the formation of photons or electron-positron pairs.
This is the ``Landau-Pomeranchuk-Midgal'' (LPM) effect.
The momentum transfer $q$ in a given event decreases with increasing incident particle energy, and so the LPM effect will suppress radiative processes for energies greater than some scale $E_\LPM$.
This can be calculated semi-classically~\cite{Klein:1998du},
\begin{align}
  E_\LPM = \frac{m_e^2 X_0 \alpha}{4 \pi}
  \approx 1~\MeV \l \frac{10^{32} \cm^{-3}}{n_\ion} \r.
\label{eq:LPM}
\end{align}
which is quite small due to the high ion density in the WD.
The stopping power for bremsstrahlung and pair production in the regime of LPM suppression $E > E_\LPM$ is
\begin{equation}
\label{eq:bremloss}
\frac{dE}{dx} \sim  \frac{E}{X_0} \l\frac{E_\LPM}{E} \r^{1/2} ~~~ E>E_\LPM.
\end{equation}
In addition to the LPM effect, soft bremsstrahlung may be suppressed in a medium as the emitted photon acquires an effective mass of order the plasma frequency $\Omega_p$.
However, for high-energy electrons this dielectric suppression only introduces a minor correction to~\eqref{eq:bremloss}, in which soft radiation is already suppressed~\cite{Klein:1998du}.

\subsection{Elastic EM Scattering}
\label{sec:coulomb}

\paragraph{Electron Coulomb Scattering off Ions.}
\label{sec:coulomb_ion}
Coulomb collisions with ions are the mechanism by which electrons of energy $1- 10~\MeV$ ultimately thermalize ions.
In this scenario we may treat the ions as stationary and ignore their recoil during collisions.
The nuclear charge will be screened by the mobile electrons of the medium, so incident particles scatter via a potential
\begin{align}
  \label{eq:ScreenedPotential}
V(\textbf{r}) = \frac{Z \alpha}{r} e^{-r/\lambda_\TF}.
\end{align}
The screening length $\lambda_\TF$ is given in the Thomas-Fermi approximation by~\cite{Teukolsky}:
\begin{align}
\label{eq:TF}
    \lambda_\TF^{2} = \frac{E_F}{6 \pi \alpha n_e}
    \sim \frac{1}{\alpha E_F^2}.
\end{align}
This plasma screening suppresses scatters with momentum transfers below $\sim \lambda_\TF^{-1}$, corresponding to a minimal energy transfer of $\omega_\xmin = \lambda_\TF^{-2} / 2 m_\ion$.
Ions may in principle also cause screening through lattice distortion, however this may be ignored as the sound speed of the lattice $c_s \sim 10^{-2}$ is much smaller than the speed of an incident relativistic electron.
From the Born approximation, the cross section for energy transfer $\omega$ is
\begin{align}
\label{eq:CoulombOffIonsCrossSection}
  \frac{d \sigma}{d \omega} =
  \frac{2 \pi Z^2 \alpha^2}{m_\ion v_\x{in}^2}
  \frac{1}{(\omega + \omega_\xmin)^2},
\end{align}
where $v_\x{in}$ is the incident velocity.
Thus the stopping power is
\begin{align}
  \frac{dE}{d x} &= \int_{0}^{\omega_\xmax} d \omega \, n_\ion
  \frac{d \sigma}{d \omega} \omega \nonumber \\
  \label{eq:StoppingPowerOffIons}
   &\approx \frac{2 \pi\, n_\ion Z^2 \alpha^2 }{m_\ion v_\x{in}^2}
   \log\left( \frac{\omega_\xmax}{\omega_\xmin} \right),
\end{align}
where the second line is valid if $\omega_\xmax \gg \omega_\xmin$.
$\omega_\xmax$ is the maximum possible energy transfer.
This may be due to 4-momentum conservation, or in the case of incident electrons, the impossibility of scattering to a final energy less than $E_F$.
4-momentum conservation sets an upper bound $\omega_\kin$, which for a stationary target is
\begin{align}
  \omega_\kin &= \frac{2 m_\ion p^2}{m_\ion^2 + m^2 + 2E m_\ion},
\end{align}
with $p$, $E$ the incoming momentum and energy.
The Fermi upper bound is $\omega_F = E - E_F$ so for incident electrons we take $\omega_\xmax = \min\left\{\omega_\kin, \omega_F\right\}$.

For scatters that transfer energy less than the plasma frequency $\Omega_p$, one may be concerned about phonon excitations.
This occurs for incident electrons with energy below $\sim 10~\MeV$.
We estimate this stopping power treating each ion as an independent oscillator with frequency $\Omega_p$ (an Einstein solid approximation) and compute the stopping power due to scatters which excite a single oscillator quanta.
There are two key differences between this and the free ion case: incident particles must transfer an energy $\Omega_p$, and the cross section to transfer momentum $q$ is suppressed by a factor $q^2 / 2 m_\ion\Omega_p = \omega_\x{free}/\Omega_p$.
$\omega_\x{free}$ is the energy transfer that would accompany a free ion scatter with momentum transfer $q$.
The resulting stopping power is unchanged from the free case~\eqref{eq:StoppingPowerOffIons}, as the increased energy transfer compensates for the suppressed cross section.

As electrons transfer their energy at the rate \eqref{eq:StoppingPowerOffIons}, they occasionally experience a hard scatter with mean free path
\begin{align}
\lambda_\text{hard} &\approx \frac{p^2 v_\text{in}^2}{\pi n_\text{ion}Z^2 \alpha^2}.
\end{align}
For sufficiently small incident energies, the electron experiences several hard scatters before it has deposited its energy by elastic scatters, and the stopping length is reduced by the resulting random walk.
This effect is not significant for incident pions due to their larger mass.

Finally, we note that for highly energetic incident particles the cross section~\eqref{eq:CoulombOffIonsCrossSection} should be modified to account for the recoil of the ion.
However, at such energies the dominant stopping power will be from hadronic or electromagnetic showers anyway, so we do not include these recoil effects.

\paragraph{Relativistic Coulomb Scattering off Electrons.}
\label{sec:coulomb_elec}
The scattering of incident electrons off degenerate electrons determines the termination energy of electromagnetic showers.
This calculation demands two considerations not present when scattering off ions: the targets are not stationary and they require a threshold energy transfer in order to be scattered out of the Fermi sea.
However for relativistic incident particle, with momentum $p \gg p_F$, the stopping power off electrons is ultimately of the same form as the stopping power off ions~\eqref{eq:StoppingPowerOffIons}.
In this limit, all particle velocities and the relative velocity is $\OO(1)$, and the deflection of the incident particle will generally be small.
It is reasonable then that scattering proceeds, up to $\OO(1)$ factors, as though a heavy incident particle is striking a light, stationary target.
The cross section is given by the usual result,
\begin{align}
  \frac{d \sigma}{d \omega} \approx
  \frac{2 \pi \alpha^2}{E_F} \frac{1}{\omega^2},
  \label{eq:CoulombRelativisticApprox}
\end{align}
where we have accounted for the target's motion by replacing its mass with its relativistic inertia $\approx E_F$.
This is equivalent to a boost of the cross section from the rest frame of the target into the WD frame.
Note that plasma screening can be ignored in this case, as Pauli-blocking will provide a more stringent cutoff on soft scatters.
Scatters which transfer an energy $\omega \leq E_F$ will have a suppressed contribution to the stopping power as they can only access a fraction of the Fermi sea.
In this limit it is sufficient to ignore these suppressed scatters:
\begin{align}
  \frac{dE}{d x} &= \int_{E_F}^{\omega_\xmax} d \omega \, n_e
  \frac{d \sigma}{d \omega} \omega \nonumber\\
  \label{eq:StoppingPowerOffElectrons}
   &\approx \frac{2 \pi\, n_e \alpha^2 }{E_F}
   \log\left( \frac{\omega_\xmax}{E_F} \right)
\end{align}
where, as described above, $\omega_\xmax = \min\{\omega_\kin, \omega_F\}$.
This derivation is admittedly quite heuristic, and so it has been checked with a detailed numerical calculation accounting fully for the target's motion and degeneracy.
Equation~\eqref{eq:StoppingPowerOffElectrons} is indeed a good approximation to the stopping power for incident energies larger than the Fermi energy.

\paragraph{Non-Relativistic Coulomb Scattering off Electrons}
For non-relativistic incident particles, the Coulomb stopping off electrons becomes strongly suppressed due to degeneracy.
Stopping in this limit appears qualitatively different than in the typical case---the slow incident particle is now bombarded by relativistic electrons from all directions.
Note that only those scatters which slow the incident particle are allowed by Pauli-blocking.

As the electron speeds are much faster than the incident, a WD electron with momentum $p_F$ will scatter to leading order with only a change in direction, so the momentum transfer is $|\vec{q}| \sim p_F$.
We again take the incident momentum $p \gtrsim p_F$, which is valid for all incident particles we consider. This results in an energy transfer
\begin{align}
\label{eq:NonRelEnergyTransfer}
  \omega = \left|\frac{p^2}{2 m} -
    \frac{\left(\vec{p} - \vec{q}\right)^2}{2 m}\right|
    \sim v_\x{in} E_F.
\end{align}
For $v_\x{in} \ll 1$ the energy transfer is less than Fermi energy, so Pauli-blocking will be important.
The incident particle is only be able to scatter from an effective electron number density
\begin{align}
  \label{eq:neff}
    n_\x{eff} = \int_{E_F - \omega}^{E_F} g(E) \; dE
    \approx 3 n_e \frac{\omega}{E_f},
\end{align}
where $g(E)$ is the Fermi density of states.
At leading order the electron is not aware of the small incident velocity, so the cross section is given by relativistic Coulomb scattering off a stationary target $\sigma \sim \alpha^2/q^2$~\cite{Jackson}.
The incident particle thus loses energy to degenerate electrons at a rate:
\begin{align}
  \frac{dE}{dt} \sim n_\x{eff} \; \sigma \; \omega
  \sim n_e \frac{\alpha^2}{E_F} v_\x{in}^2.
\end{align}
Note that this includes a factor of the relative velocity which is $\OO(1)$.
As a result, the stopping power is parametrically
\begin{align}
  \label{eq:degnonrel}
  \frac{dE}{dx} =  \frac{1}{v_\x{in}} \frac{dE}{dt} \sim
  n_e \frac{\alpha^2}{E_F} v_\x{in}.
\end{align}
As above, this heuristic result has been verified with a full integration of the relativistic cross section.

We can compare~\eqref{eq:degnonrel} to the stopping power of non-relativistic, heavy particles off roughly stationary, non-degenerate electrons $\frac{dE}{dx} \sim n_e \frac{\alpha^2}{m_e v_\text{in}^2}$, which is the familiar setting of stopping charged particles in a solid due to ionization~\cite{Rossi}.
Evidently, the analogous stopping in a WD is parametrically suppressed by $v_\x{in}^3 m_e/E_F$.
One factor of $v_\x{in}$ is due to Pauli blocking, while the other factors are kinematic, due to the relativistic motion of the targets.

\paragraph{Compton Scattering}
\label{sec:compton}
Compton scattering off degenerate electrons is the dominant interaction for photons of incident energy $k \leq E_F$.
As we will show, this stopping power is parametrically different from that of high-energy photons due to Pauli-blocking and the motion of the electron.
For $k>E_F$, the effect of Pauli-blocking is negligible and the stopping power is simply:
\begin{equation}
\frac{dk}{dx} \sim \frac{\pi \alpha^2 n_e}{E_F} \log\l \frac{k}{m_e}\r,
\end{equation}
where again we have (partially) applied the heuristic $m_e \rightarrow E_F$ replacement to boost the usual result for stationary electrons while avoiding divergence at the Fermi energy.
This, along with the low-energy estimate below, matches a full integration of the relativistic cross section well.

We now turn to the regime of interest, $k < E_F$.
Only those electrons near the top of the Fermi sea are available to scatter, so the photon interacts with only the effective electron density~\eqref{eq:neff}.
In addition, Compton scatters will only occur off electrons moving roughly collinear with the photon momentum - a head-on collision would result in an energy loss for the electron, which is forbidden by Pauli exclusion.
In the electron rest frame these collinear scatters are Thompson-like, and the photon energy loss is dominated by backward scatters.
For relativistic electrons near the Fermi surface, these scatters transfer an energy
\begin{align}
  \omega \sim k \l 1 - \frac{m_e^2}{4 E_F^2} \r \approx k.
\end{align}
The cross section can be taken in the electron rest frame $\sigma \sim \alpha^2/m_e^2$, along with an `aiming' factor $1/4\pi$ to account for the restriction to initially parallel trajectories.
This gives a stopping power
\begin{align}
  \frac{dk}{dx} \approx \frac{\alpha^2 n_e k^2}{4 \pi m_e^2 E_F}.
\end{align}

\section{Dark Matter Capture}
\label{sec:capture}
Here we give a more detailed discussion of DM capture in a WD and its subsequent evolution.
For the remainder of this section all numerical quantities are evaluated at a central WD density $\rho_\text{WD} \sim 3 \times 10^{8} \frac{\text{g}}{\cm^{3}}$ ($n_\text{ion} \sim 10^{31} ~\cm^{-3}$), for which the relevant WD parameters are~\cite{cococubed}:~$M_\text{WD} \approx 1.25 ~M_{\astrosun}$, $R_\text{WD} \approx 4000 ~\text{km}$, and~$v_\text{esc} \approx 2 \times 10^{-2}$.
Depending on the context, the relevant density may be the average value which we take to be~$\sim 10^{30} ~\cm^{-3}$.
We also assume an average value of the WD temperature $T_\text{WD} \sim \text{keV}$.

\subsection{Capture Rate}
Consider spin-independent DM elastic scattering off ions with cross section $\sigma_{\chi A}$.
This is related to the per-nucleon cross section
\begin{equation}
\sigma_{\chi A} = A^2 \l \frac{\mu_{\chi A}}{\mu_{\chi n}}\r^2 F^2(q) \sigma_{\chi n} = A^4 F^2(q) \sigma_{\chi n},
\end{equation}
where $F^2(q)$ is the Helm form factor~\cite{Helm:1956zz}.
If the DM is at the WD escape velocity, the typical momentum transfer to ions is $q \sim \mu_{\chi A} v_\text{esc} \sim 200 ~\MeV$.
As this $q$ is less than or of order the inverse nuclear size, DM scattering off nuclei will be coherently enhanced.
We find $F^2(q) \approx 0.1$ for $q \sim 200 ~\MeV$.

For the DM to ultimately be captured, it must lose energy $\sim m_\chi v^2$, where $v$ is the DM velocity (in the rest frame of the WD) asymptotically far away.
Since typically $v \ll v_\text{esc}$, the DM has velocity $v_\text{esc}$ while in the star and must lose a fraction $(v/v_\text{esc})^2$ of its kinetic energy to become captured.
Properly, the DM velocity is described by a boosted Maxwell distribution peaked at the galactic virial velocity $v_\text{halo} \sim 10^{-3}$.
However, this differs from the ordinary Maxwell distribution by only $\OO(1)$ factors~\cite{Gould:1987ir}, and we can approximate it by (ignoring the exponential Boltzmann tail):
\begin{equation}
\frac{dn_\chi}{dv} \approx
\begin{cases}
  \frac{\rho_\chi}{m_\chi} \l \frac{v^2}{v_\text{halo}^3} \r  & v \leq v_\text{halo} \\
  0 & v > v_\text{halo}
  \end{cases}.
\end{equation}
The DM capture rate is given by an integral of the DM transit rate weighted by a probability for capture $P_\text{cap}$
\begin{equation}
\Gamma_\text{cap} \sim \int dv \frac{d \Gamma_\text{trans}}{dv} P_\text{cap}(v),
\end{equation}
where the (differential) transit rate is
\begin{equation}
\frac{d \Gamma_\text{trans}}{dv} \sim \frac{d n_\chi}{dv} R_\text{WD}^2 \l \frac{v_\text{esc}}{v}\r^2 v.
\end{equation}
$P_\text{cap}$ depends on both the \emph{average} number of scatters in a WD
\begin{equation}
\overbar{N}_\text{scat} \sim n_\text{ion} \sigma_{\chi A} R_\text{WD},
\end{equation}
and the number of scatters \emph{needed} for capture
\begin{equation}
N_\text{cap} \sim \text{max}\left \{1, \frac{m_\chi v^2}{m_\text{ion} v_\text{esc}^2}\right \},
\end{equation}
and is most generally expressed as a Poisson sum
\begin{equation}
P_\text{cap} = 1 - \sum^{N_\text{cap}-1}_{n=0} \exp(-\overbar{N}_\text{scat})\frac{(\overbar{N}_\text{scat})^n}{n!}.
\end{equation}
For our purposes we will approximate the sum as follows:
\begin{equation}
P_\text{cap} \approx
\begin{cases}
 1 & \overbar{N}_\text{scat} > N_\text{cap} \\
 \overbar{N}_\text{scat} & \overbar{N}_\text{scat} < N_\text{cap} ~\text{and}~ N_\text{cap} = 1 \\
 0 & \text{else}
\end{cases}.
\end{equation}
Here we ignore the possibly of capture if $\overbar{N}_\text{scat} < N_\text{cap}$ except in the special case that only one scatter is needed for capture.
If $\overbar{N}_\text{scat} > N_\text{cap}$, we assume all DM is captured.
Most accurately, this capture rate should be computed numerically, e.g. see~\cite{Bramante:2017xlb}.
However with the above simplifications we find that the capture rate is of order
\begin{align}
  \Gamma_\text{cap} &\sim \Gamma_\text{trans} \cdot
  \text{min}\left\{1, \overbar{N}_\text{scat} \text{min}\{B,1\}\right\}, \\
  B &\equiv \frac{m_\text{ion} v_\text{esc}^2}{m_\chi v_\text{halo}^2}.
  \nonumber
\end{align}
$B$ here encodes the necessity of multiple scattering for capture.
For ultra-heavy DM $m_\chi > 10^{15} ~\GeV$, $B \ll 1$ and essentially multiple scatters are always needed.

\subsection{Thermalization and Collapse}

Once DM is captured, it thermalizes to an average velocity
\begin{equation}
  v_\text{th} \sim \sqrt{\frac{T_\text{WD}}{m_\chi}}
  \approx 10^{-11} \l \frac{m_\chi}{10^{16} ~\GeV}\r^{-1/2},
\end{equation}
and settles to the thermal radius
\begin{align}
  R_\text{th} \sim \l \frac{T_\text{WD}}{G m_\chi \rho_\text{WD}}\r^{1/2}
 \approx 0.1 ~\cm \l \frac{m_\chi}{10^{16} ~\GeV}\r^{-1/2}, \nonumber
\end{align}
where its kinetic energy balances against the gravitational potential energy of the (enclosed) WD mass.
This thermalization time can be explicitly calculated for elastic nuclear scatters~\cite{Kouvaris:2010jy}.
The stopping power due to such scatters is
\begin{align}
    \frac{dE}{dx} \sim \rho_\text{WD} \sigma_{\chi A} \; v \; \text{max}\{v, v_\ion\},
\end{align}
where $v_\ion \sim \sqrt{T_\text{WD}/m_\text{ion}}$ is the thermal ion velocity.
The max function indicates the transition between ``inertial" and ``viscous" drag, as the DM velocity $v$ slows to below $v_\text{ion}$.
DM first passes through the WD many times on a wide orbit until the size of its orbit decays to become contained in the star.
The timescale for this process is
\begin{align}
  t_1 &\sim \l \frac{m_\chi}{m_\text{ion}} \r^{3/2}
  \frac{R_\text{WD}}{v_\text{esc}} \frac{1}{\overbar{N}_\text{scat}}
  \frac{1}{\text{max}\{\overbar{N}_\text{scat}, 1\}^{1/2}} \\
  &\approx 7 \times 10^{16}~\text{s} \l \frac{m_\chi}{10^{16} ~\GeV} \r^{3/2}
  \l \frac{\sigma_{\chi A}}{10^{-35} ~\cm^2} \r^{-3/2}. \nonumber
\end{align}
Subsequently, the DM completes many orbits within the star until dissipation further reduces the orbital size to the thermal radius.
The timescale for this process is
\begin{align}
  t_2  &\sim \l \frac{m_\chi}{m_\text{ion}} \r
  \frac{1}{n_\text{ion} \sigma_{\chi A}} \frac{1}{v_\text{ion}} \\
  &\approx 10^{14}~\text{s}\l \frac{m_\chi}{10^{16} ~\GeV} \r
  \l \frac{\sigma_{\chi A}}{10^{-35} ~\cm^2} \r^{-1}. \nonumber
\end{align}
There is an additional $\OO(10)$ logarithmic enhancement of the timescale once the DM velocity has slowed below $v_\ion$.
Note that time to complete a single orbit is set by the gravitational free-fall timescale:
\begin{equation}
\label{eq:freefalltime}
t_\text{ff} \sim \sqrt{\frac{1}{G \rho_\text{WD}}} \approx 0.5 ~\text{s}.
\end{equation}

In the above description, we have assumed that the DM loses a negligible amount of energy during a single transit:
\begin{equation}
\frac{\sigma_{\chi A}}{m_\chi} \ll \frac{1}{\rho_\text{WD} R_\text{WD}}.
\end{equation}
This also ensures that the dynamics of DM within the star is that of Newtonian gravity along with a small drag force.
In the opposite regime, the qualitative evolution of captured DM differs from the picture presented in detail below.
In this case there is no stage of external orbital motion corresponding to $t_1$---DM will instead rapidly thermalize to a speed $v_\x{th}$ after entering the star.
The internal motion now proceeds as a gravitationally-biased random walk, with a net drift of DM towards the center of the star.
For sufficiently large $\sigma_{\chi A}$, DM will collect at a radius $r_\x{c}$ which is larger than $r_\x{th}$ given above, due to a balance of gravity with outward Brownian diffusion.
This may delay the onset of self-gravitation, possibly beyond $\tau_\x{WD}$, as we now require the collection of a larger mass $\rho_\x{WD} r^3_\x{c}$.
It is important to note that the differences between the Brownian and orbital regimes are immaterial for constraints on the decay of captured DM (e.g., Figure~\ref{fig:capture-decay}), which cares only about the quantity of DM present in the star.
For annihilation constraints, however, the internal evolution of DM is quite important.
For the largest unconstrained cross sections $\sigma_{\chi A}$ (see Figure~\ref{fig:elastic-capture}), one can check that captured DM is distributed across a large fraction of the star due to Brownian motion and does not collapse.
This DM population still yields a strong constraint on $\sigma_{\chi \chi}$, similar to but somewhat weaker than the constraints which can be placed on DM that undergoes self-gravitational collapse after capture (e.g., Figure~\ref{fig:capture-collision}).

When Brownian motion is insignificant, the DM will begin steadily accumulating at $R_\text{th}$ after a time $t_1 + t_2$.
Once the collected mass of DM at the thermal radius exceeds the WD mass within this volume, there is the possibility of self-gravitational collapse.
The time to collect a critical number $N_\text{sg}$ of DM particles is
\begin{align}
\label{eq:Ncore}
    t_\text{sg} &\sim \frac{N_\text{sg}}{\Gamma_\text{cap}}  \sim
    \frac{\rho_\text{WD} R^3_\text{th}}{m_\chi \Gamma_\text{cap}} \\
    &\approx 10^{10} ~\text{s} \l \frac{m_\chi}{10^{16} ~\GeV} \r^{-1/2}
    \l \frac{\sigma_{\chi A}}{10^{-35} ~\cm^2} \r^{-1}, \nonumber
\end{align}
Typically, the timescale for collapse is then set by the DM sphere's ability to cool and shed gravitational potential energy.
This is initially just $t_2$, while the time to collapse at any given radius $r$ decreases once the DM velocity rises again above $v_\ion$:
\begin{align}
  t_\text{cool} &\sim t_2 \text{min}\{v_\text{ion}/v_\chi,1\} \\
  v_\chi &\sim \sqrt{\frac{G N m_\chi}{r}}, \nonumber
\end{align}
where $N$ is the number of collapsing DM particles.
Note that when $m_\chi > 10^{21} ~\GeV$, the number of particles necessary for self-gravitation $N_\text{sg}$ as defined in \eqref{eq:Ncore} is less than $2$.
In this case we should formally take $N_\text{sg} = 2$.

Finally, there is a further subtlety that arises in the growing of DM cores for the large DM masses $m_\chi$ of interest to us.
The time $t_\text{sg}$ to collect a self-gravitating number of particles decreases for larger DM masses.
However, the dynamics of the collapse are set by the cooling time, which is initially $t_\text{cool} \propto m_\chi$.
For $m_\chi > 10^{15} ~\GeV$, the collection time may be shorter than the cooling time $t_\text{sg} < t_\text{cool}$ (depending on the cross section).
In fact, the collection time may even be shorter than the dynamical time $t_\text{ff}$.
If $t_\text{ff} < t_\text{sg} <t_\text{cool}$, the DM core will be driven to shrink because of the gravitational potential of the over-collecting DM.
The timescale for the shrinking is set by the capture rate of DM.
Ultimately, the collapsing DM core will consist of $N_\text{sg}$ enveloped in a ``halo" of $\Gamma_\text{cap} t_\text{cool} \gg N_\text{sg}$ particles, which will also proceed to collapse.
If instead $t_\text{sg} < t_\text{ff} <t_\text{cool}$, the DM core will rapidly accumulate to this large number before dynamically adjusting.
For the purpose of the collapse constraints on DM annihilation, if $t_\x{sg} < t_\text{cool}$ we will simply assume a number of collapsing particles $N = \Gamma_\text{cap} t_\text{cool}$.
This is the case for the constraints plotted in Figure~\ref{fig:capture-collision}.
\end{appendices}

\section*{Acknowledgements}
We would like to thank Kim Berghaus, Kyle Boone, Jeff Dror, Keisuke Harigaya, David E. Kaplan, Spencer Klein, Chris Kouvaris, Jacob Leedom, Junsong Lin, Chung-Pei Ma, Sam McDermott, Katelin Schutz, Peter Tinyakov, and Lian-Tao Wang for stimulating discussions.
PWG was supported by NSF grant PHY-1720397, DOE Early Career Award DE-SC0012012, and Heising-Simons Foundation grant 2015-037.
SR was supported in part by the NSF under grants PHY-1638509 and PHY-1507160, the Alfred P. Sloan Foundation grant FG-2016-6193 and the Simons Foundation Award 378243. 



\begin{thebibliography}{99}
\bibliographystyle{unsrt}
\bibitem{Akerib:2016vxi} 
  D.~S.~Akerib {\it et al.} [LUX Collaboration],
  Phys.\ Rev.\ Lett.\  {\bf 118}, no. 2, 021303 (2017)
  [arXiv:1608.07648 [astro-ph.CO]].


\bibitem{Agnese:2017njq} 
  R.~Agnese {\it et al.} [SuperCDMS Collaboration],
  Phys.\ Rev.\ Lett.\  {\bf 120}, no. 6, 061802 (2018)
  [arXiv:1708.08869 [hep-ex]].

\bibitem{Griest:2013aaa} 
  K.~Griest, A.~M.~Cieplak and M.~J.~Lehner,
  Astrophys.\ J.\  {\bf 786}, no. 2, 158 (2014)
  [arXiv:1307.5798 [astro-ph.CO]].
  

\bibitem{Graham:2015apa} 
  P.~W.~Graham, S.~Rajendran and J.~Varela,
  Phys.\ Rev.\ D {\bf 92}, no. 6, 063007 (2015)
  [arXiv:1505.04444 [hep-ph]].


\bibitem{Maoz:2012}
  D. Maoz and F. Mannucci,
  PASA, 29, 447 (2012)
  [arXiv:1111.4492 [astro-ph.CO]].


\bibitem{Scalzo:2014sap} 
  R.~Scalzo {\it et al.} [Nearby Supernova Factory Collaboration],
  Mon.\ Not.\ Roy.\ Astron.\ Soc.\  {\bf 440}, no. 2, 1498 (2014)
  [arXiv:1402.6842 [astro-ph.CO]].


\bibitem{Scalzo:2014wxa} 
  R.~A.~Scalzo, A.~J.~Ruiter and S.~A.~Sim,
  Mon.\ Not.\ Roy.\ Astron.\ Soc.\  {\bf 445}, no. 3, 2535 (2014)
  [arXiv:1408.6601 [astro-ph.HE]].


\bibitem{McGee:2010} 
  S.~L. McGee and M.~L Balogh,
  Mon.\ Not.\ Roy.\ Astron.\ Soc.\  {\bf 403}, L79 (2010)
  [arXiv:0912.3455 [astro-ph.CO]].

\bibitem{Foley:2013}
  R.J.~Foley, P.~J. Challis, R. Chornock, et al.
  ApJ, 767, 57 (2013)
  [arXiv:1212.2209 [astro-ph.CO]].

\bibitem{Kasliwal:2012}
  M.~M. Kasliwal, S.~R. Kulkarni, A. Gal-Yam, et al.
  ApJ, 755, 161 (2012)
  [arXiv:1111.6109 [astro-ph.HE]].

\bibitem{Woosley1994}
  S.~E.~Woosley and T.~A.~Weaver, Astrophysical Journal {\bf 423}, pp.371-379 (1994).

\bibitem{Fink:2007fv} 
  M.~Fink, W.~Hillebrandt and F.~K.~Roepke,
  Astron.\ Astrophys.\ 
  [Astron.\ Astrophys.\  {\bf 476}, 1133 (2007)]
  [arXiv:0710.5486 [astro-ph]].


\bibitem{Pakmor:2013wia} 
  R.~Pakmor, M.~Kromer and S.~Taubenberger,
  Astrophys.\ J.\  {\bf 770}, L8 (2013)
  [arXiv:1302.2913 [astro-ph.HE]].


\bibitem{Sell:2015rfa} 
  P.~H.~Sell, T.~J.~Maccarone, R.~Kotak, C.~Knigge and D.~J.~Sand,
  Mon.\ Not.\ Roy.\ Astron.\ Soc.\  {\bf 450}, no. 4, 4198 (2015)
  [arXiv:1504.05584 [astro-ph.HE]].


\bibitem{Woosley}
 F.~X. Timmes and S.~E. Woosley, 
 Astro. Phys. Journal {\bf 396}, 649 (1992).
 
\bibitem{Bertone:2007ae} 
  G.~Bertone and M.~Fairbairn,
  Phys.\ Rev.\ D {\bf 77}, 043515 (2008)
  [arXiv:0709.1485 [astro-ph]].

\bibitem{McCullough:2010ai} 
  M.~McCullough and M.~Fairbairn,
  Phys.\ Rev.\ D {\bf 81}, 083520 (2010)
  [arXiv:1001.2737 [hep-ph]].

\bibitem{Leung:2013pra} 
  S.-C.~Leung, M.-C.~Chu, L.-M.~Lin and K.-W.~Wong,
  Phys.\ Rev.\ D {\bf 87}, no. 12, 123506 (2013)
  [arXiv:1305.6142 [astro-ph.CO]].

\bibitem{Bramante:2015cua} 
  J.~Bramante,
  Phys.\ Rev.\ Lett.\  {\bf 115}, no. 14, 141301 (2015)
  [arXiv:1505.07464 [hep-ph]].

\bibitem{Gasques:2005ar} 
  L.~R.~Gasques, A.~V.~Afanasjev, E.~F.~Aguilera, M.~Beard, L.~C.~Chamon, P.~Ring, M.~Wiescher and D.~G.~Yakovlev,
  Phys.\ Rev.\ C {\bf 72}, 025806 (2005)
  [astro-ph/0506386].


\bibitem{cococubed}
F.~X.~Timmes, \href{http://cococubed.asu.edu/code_pages/coldwd.shtml}{link}

\bibitem{Gandhi:1998ri} 
  R.~Gandhi, C.~Quigg, M.~H.~Reno and I.~Sarcevic,
  Phys.\ Rev.\ D {\bf 58}, 093009 (1998)
  [hep-ph/9807264].


\bibitem{Formaggio:2013kya} 
  J.~A.~Formaggio and G.~P.~Zeller,
  Rev.\ Mod.\ Phys.\  {\bf 84}, 1307 (2012)
  [arXiv:1305.7513 [hep-ex]].


\bibitem{Press:1985ug} 
  W.~H.~Press and D.~N.~Spergel,
  Astrophys.\ J.\  {\bf 296}, 679 (1985).


\bibitem{Gould:1987ir} 
  A.~Gould,
  Astrophys.\ J.\  {\bf 321}, 571 (1987).

\bibitem{us}
R.~Janish, V.~Narayan, and P.~Riggins, in preparation.

\bibitem{Mereghetti:2013nba} 
  S.~Mereghetti,
  arXiv:1302.4634 [astro-ph.HE].

\bibitem{Winget:1987}
D.~E.~Winget, C.~J.~Hansen, James Liebert {\it et al.}, 
  Astrophys.\ J.\  {\bf 315}, L77 (1987).

\bibitem{SDSS}
S.~J.~Kleinman, S. O. Kepler, D. Koester, I. Pelisoli  {\it et al.}, Astrophys. J. Suppl. {\bf 204}, article
id. 5, 14 pp. (2013)

\bibitem{NuStar}
K.~Perez, C.~J.~Hailey, F.~E.~Bauer, {\it et al.}, Nature {\bf 520}, 646 (2015)

\bibitem{Nesti:2013uwa} 
  F.~Nesti and P.~Salucci,
  JCAP {\bf 1307}, 016 (2013)
  [arXiv:1304.5127 [astro-ph.GA]].


\bibitem{Chandrasekhar}
S.~Chandrasekhar, ``An Introduction to the Study of Stellar Structure", University of Chicago press (1939).

\bibitem{KippenhahnWeigert}
R.~Kippenhahn and A.~Weigert, ``Stellar Structure and Evolution", Springer (1994).

\bibitem{Dvorkin:2013cea} 
  C.~Dvorkin, K.~Blum and M.~Kamionkowski,
  Phys.\ Rev.\ D {\bf 89}, no. 2, 023519 (2014)
  [arXiv:1311.2937 [astro-ph.CO]].


\bibitem{Padmanabhan:2005es} 
  N.~Padmanabhan and D.~P.~Finkbeiner,
  Phys.\ Rev.\ D {\bf 72}, 023508 (2005)
  [astro-ph/0503486].


\bibitem{Slatyer:2009yq} 
  T.~R.~Slatyer, N.~Padmanabhan and D.~P.~Finkbeiner,
  Phys.\ Rev.\ D {\bf 80}, 043526 (2009)
  [arXiv:0906.1197 [astro-ph.CO]].


\bibitem{Slatyer:2016qyl} 
  T.~R.~Slatyer and C.~L.~Wu,
  Phys.\ Rev.\ D {\bf 95}, no. 2, 023010 (2017)
  [arXiv:1610.06933 [astro-ph.CO]].


\bibitem{ThePierreAuger:2015rma} 
  A.~Aab {\it et al.} [Pierre Auger Collaboration],
  Nucl.\ Instrum.\ Meth.\ A {\bf 798}, 172 (2015)
  [arXiv:1502.01323 [astro-ph.IM]].


\bibitem{Mack:2007xj} 
  G.~D.~Mack, J.~F.~Beacom and G.~Bertone,
  Phys.\ Rev.\ D {\bf 76}, 043523 (2007)
  [arXiv:0705.4298 [astro-ph]].


\bibitem{Aprile:2017iyp} 
  E.~Aprile {\it et al.} [XENON Collaboration],
  Phys.\ Rev.\ Lett.\  {\bf 119}, no. 18, 181301 (2017)
  [arXiv:1705.06655 [astro-ph.CO]].


\bibitem{Ambrosio:2002qq} 
  M.~Ambrosio {\it et al.} [MACRO Collaboration],
  Eur.\ Phys.\ J.\ C {\bf 25}, 511 (2002)
  [hep-ex/0207020].


\bibitem{Jacobs:2014yca} 
  D.~M.~Jacobs, G.~D.~Starkman and B.~W.~Lynn,
  Mon.\ Not.\ Roy.\ Astron.\ Soc.\  {\bf 450}, no. 4, 3418 (2015)
  [arXiv:1410.2236 [astro-ph.CO]].


\bibitem{Coleman:1985ki} 
  S.~R.~Coleman,
  Nucl.\ Phys.\ B {\bf 262}, 263 (1985)
  Erratum: [Nucl.\ Phys.\ B {\bf 269}, 744 (1986)].


\bibitem{Kusenko:1997si} 
  A.~Kusenko and M.~E.~Shaposhnikov,
  Phys.\ Lett.\ B {\bf 418}, 46 (1998)
  [hep-ph/9709492].


\bibitem{Kusenko:1997vp} 
  A.~Kusenko, V.~Kuzmin, M.~E.~Shaposhnikov and P.~G.~Tinyakov,
  Phys.\ Rev.\ Lett.\  {\bf 80}, 3185 (1998)
  [hep-ph/9712212].

\bibitem{Dine:2003ax} 
  M.~Dine and A.~Kusenko,
  Rev.\ Mod.\ Phys.\  {\bf 76}, 1 (2003)
  [hep-ph/0303065].


\bibitem{Tavernier}
S.~Tavernier, ``Experimental Techniques in Nuclear and Particle Physics", Springer (2010).

\bibitem{Pionnuclear}
T.~S.~H.~Lee and R.~P.~Redwine,
 Annu. Rev. Nucl. Part. Sci {\bf 52}, pp.23-63 (2002)


\bibitem{Gerhardt:2010bj} 
  L.~Gerhardt and S.~R.~Klein,
  Phys.\ Rev.\ D {\bf 82}, 074017 (2010)
  [arXiv:1007.0039 [hep-ph]].


\bibitem{Klein:1998du} 
  S.~Klein,
  Rev.\ Mod.\ Phys.\  {\bf 71}, 1501 (1999)
  [hep-ph/9802442].


\bibitem{Bethe1934}
  H.~Bethe and W.~Heitler
  Proc.\ R.\ Soc.\ Lond.\ A 1934 146 83-112

\bibitem{Teukolsky}
S.~L.~Shapiro and S.~A.~Teukolsky, ``Black Holes, White Dwarfs, and Neutron Stars", Wiley (1983).

\bibitem{Jackson}
J.~D.~Jackson, ``Classical Electrodynamics", 3rd edition, John Wiley and Sons, New
York, (1998).

\bibitem{Rossi}
B.~Rossi, ``High Energy Particles", Prentice-Hall, Inc., Englewood Cliffs, NJ (1952).

\bibitem{Helm:1956zz} 
  R.~H.~Helm,
  Phys.\ Rev.\  {\bf 104}, 1466 (1956).


\bibitem{Bramante:2017xlb} 
  J.~Bramante, A.~Delgado and A.~Martin,
  Phys.\ Rev.\ D {\bf 96}, no. 6, 063002 (2017)
  [arXiv:1703.04043 [hep-ph]].


\bibitem{Kouvaris:2010jy} 
  C.~Kouvaris and P.~Tinyakov,
  Phys.\ Rev.\ D {\bf 83}, 083512 (2011)
  [arXiv:1012.2039 [astro-ph.HE]].

\end{thebibliography}
\end{document}